\pdfoutput=1
\documentclass[11pt,a4paper]{article}
\usepackage{jheppub}
\usepackage{amsmath,amssymb}
\usepackage[caption=false]{subfig}
\usepackage{graphicx}
\usepackage{hyperref}
\usepackage{multirow}
\usepackage{color}
\usepackage[figuresleft]{rotating}

\newcommand{\Lag}{\mathcal{L}}
\newcommand{\p}{\partial}

\newcommand{\beq}{\begin{eqnarray}}
\newcommand{\eeq}{\end{eqnarray}}
\newcommand{\non}{\nonumber\\}

\newcommand{\Tr}{{\rm Tr}\,}

\newcommand{\btau}{\boldsymbol{\tau}}
\newcommand{\bpi}{\boldsymbol{\pi}}

\begin{document}

\title{Exploring the generalized loosely bound Skyrme model}

\author{Sven Bjarke Gudnason}
\affiliation{Department of Physics, and Research and Education
  Center for Natural Sciences, Keio University, Hiyoshi 4-1-1,
  Yokohama, Kanagawa 223-8521, Japan}
\emailAdd{gudnason(at)keio.jp}

\abstract{
The Skyrme model is extended with a sextic derivative term -- called
the BPS-Skyrme term -- and a repulsive potential term --  called the
loosely bound potential.
A large part of the model's parameter space is studied for the
4-Skyrmion which corresponds to the Helium-4 nucleus and emphasis is
put on preserving as much of the platonic symmetries as possible
whilst reducing the binding energies. 
We reach classical binding energies for Helium-4 as low as $0.2\%$,
while retaining the cubic symmetry of the 4-Skyrmion, and after taking
into account the quantum mass correction to the nucleon due to
spin/isospin quantization, we get total binding energies as low as
$3.6\%$ -- still with the cubic symmetry intact.
}


\maketitle

\section{Introduction}

The Skyrme model is an interesting approach to nuclear physics which
is related to fundamental physics, i.e.~it is a field
theory \cite{Skyrme:1962vh,Skyrme:1961vq}; indeed the soliton in the
model is identified with the baryon in the large-$N$ limit of
QCD \cite{Witten:1983tw,Witten:1983tx}.
The soliton is called the Skyrmion.
The most interesting aspect of the Skyrmions is that although a single
Skyrmion is identified with a single nucleon, multi-Skyrmion solutions
exist which can be identified with nuclei of higher baryon numbers.
This fact distinguishes the Skyrme model from basically all other
approaches to nuclear physics; the nuclei are no longer bound states of
interacting point particles.
What is more interesting is that since even the single Skyrmion is a
spatially extended object (as opposed to a point particle), the
multi-Skyrmions become extended objects with certain platonic
symmetries \cite{Braaten:1989rg}.
A particular useful Ansatz for light nuclei was found using a rational
map \cite{Houghton:1997kg}.
For larger nuclei, however, there is some consensus that when a pion
mass term is included in the model, they are made of $B/4$ cubes --
akin somewhat to the alpha particle model of 
nuclei \cite{Battye:2006na,Feist:2012ps,Lau:2014baa,Halcrow:2016spb,Rawlinson:2017rcq} 
-- up to small deformations.

A long standing problem of the Skyrme model as a model for nuclei is
that the multi-Skyrmions are too strongly bound; their binding
energies are about one order of magnitude larger than what is
measured in nuclei experimentally.
One approach to solving this problem is based on modifying the Skyrme 
model such that the classical solutions can come close to a BPS-like 
energy bound, in which case the classical energy is approximately
proportional to the topological charge.
Hence, if such a bound could be saturated, then classically the
binding energy would vanish exactly.
In the last decade, this approach has been taken in three different
directions: the vector meson Skyrme
model \cite{Sutcliffe:2010et,Sutcliffe:2011ig,Naya:2018mpt}, the
BPS-Skyrme model \cite{Adam:2010fg,Adam:2010ds} and 
the lightly/loosely bound Skyrme
models \cite{Harland:2013rxa,Gillard:2015eia,Gudnason:2016mms}. 

The vector meson Skyrme model is inspired by approximate Skyrmion
solutions obtained from instanton holonomies \cite{Atiyah:1992if} and
it is found that the instanton holonomy becomes an exact solution in
the limit where an infinite tower of vector mesons is included; the 
complete theory can be described simply as a 4+1 dimensional
Yang-Mills (YM) theory in flat spacetime \cite{Sutcliffe:2010et}. 
The instanton is a half-BPS state in the YM theory and the Skyrmion
saturates the BPS bound if the theory is not truncated.
The standard Skyrme model is at the other end of the scale; all vector
mesons have been stripped off, leaving behind just the pions.
The model is in very close relation to holography, although the
discrete spectrum of vector mesons is due to truncation of the theory
and not due to an intrinsic curvature of the background
spacetime \cite{Sutcliffe:2010et}. 
Indeed in the Sakai-Sugimoto model, the standard Skyrme model comes
out as the low-energy action of the zero modes \cite{Sakai:2004cn}.
The sextic term, which we shall discuss shortly in a different
context, also comes out naturally by integrating out the first vector
meson in the Sakai-Sugimoto model \cite{Bartolini:2017sxi}.

The BPS-Skyrme model is based on a drastic modification of the Skyrme
model: remove the original terms and replace them with a sextic term,
which we shall call the BPS-Skyrme term, and a
potential \cite{Adam:2010fg,Adam:2010ds}.
The model has the advantage of simplicity in the following sense: the
model in this limit is not only a BPS theory -- in the sense of
being able to saturate a BPS-like energy bound -- but it is also
integrable.
That is, a large class of exact analytic solutions has been
obtained.
The disadvantage is that the kinetic term and the Skyrme term have to
be quite suppressed in order for the classical binding energy to
be of the order of magnitude seen in experimental data.
That issue is two-fold; the first problem is that we would like to
maintain the coefficient of the kinetic term (the pion decay constant)
and pion mass of the order of the measured values in the pion vacuum. 
The second problem is of a more technical nature; close to the BPS
limit, the coefficient of the kinetic term ($c_2$) is very small and
hence there are certain points/lines in the Skyrmion solutions where
the solution can afford to have very large field derivatives -- of the
order of $1/\sqrt{c_2}$.
In ref.~\cite{Gillard:2015eia} the order of magnitude of $c_2$ for the
classical binding energies to be in the ballpark of the experimental
values was estimated to be around $c_2\sim 0.01$; whereas their
numerics was trustable only down to about $c_2\sim 0.2$.

The lightly bound Skyrme model is based on an energy
bound \cite{Harland:2013rxa,Adam:2013tga} for the Skyrme term and a
potential to the fourth power.
Although this model has a saturable solution in the 1-Skyrmion sector,
no solutions saturate the bound for higher topological degrees.
Nevertheless, it turns out that, although the solutions do not saturate
the bound for higher topological degrees, they can come quite close to
the bound, which in turn yields a small classical binding
energy \cite{Gillard:2015eia}.
The lightly bound model can indeed reduce the binding energies, but it
comes with a price; long before realistic binding energies are
reached, the platonic symmetries of the compact Skyrmions are lost and
the potential has the effect of pushing out identifiable 1-Skyrmions
which remain only very weakly bound.
This limit was the inspiration for a simplified kind of Skyrme model,
called the point particle model of lightly bound
Skyrmions \cite{Gillard:2016esy}.
Said limit can also be obtained naturally in the Sakai-Sugimoto model
by considering the strong 't Hooft coupling
limit \cite{Baldino:2017mqq}.

In ref.~\cite{Gudnason:2016mms}, we compared the potential made of the
standard pion mass term to the fourth power and the same potential to
the second power; we call them the lightly bound and the loosely bound
potentials, respectively.
It turns out that the loosely bound potential has the same repulsive
effect as the lightly bound potential does, but the Skyrmions retain
their platonic symmetries down to smaller binding energies for the
loosely bound potential as compared to the lightly bound one. 
In ref.~\cite{Gudnason:2016cdo}, we further established that if the
potential is treated as a polynomial in $\sigma=\tfrac{1}{2}\Tr[U]$,
where $U$ is the chiral field, then to second order, the
loosely bound potential is the potential that can reduce the binding
energy the most while preserving platonic symmetries of the
Skyrmions.

In ref.~\cite{Gudnason:2016tiz}, we expanded the model by making a
hybrid model out of the BPS-Skyrme-type models and the loosely bound
model.
Let us define the generalized Skyrme model as the standard Skyrme
model with the addition of the sextic BPS-Skyrme term. 
Thus in ref.~\cite{Gudnason:2016tiz} we studied the generalized Skyrme
model with the pion mass term and the loosely bound potential.
More precisely, we studied the model using the rational map Ansatz for
the 4-Skyrmion in the regime where the coefficients for the BPS-Skyrme
term, $c_6$, and the loosely bound potential, $m_2$, were both taken
to be small (i.e.~smaller or equal to one). 
Physical effects on the observables of the model could readily be
extracted without the effort of full numerical PDE (partial
differential equation) calculations.
Of course, that compromise had the consequence that we could not
detect the change of symmetry in the Skyrmion solutions and hence we
investigated only a very restricted part of the parameter space.

\begin{figure}[!htp]
\begin{center}
\includegraphics[width=0.6\linewidth]{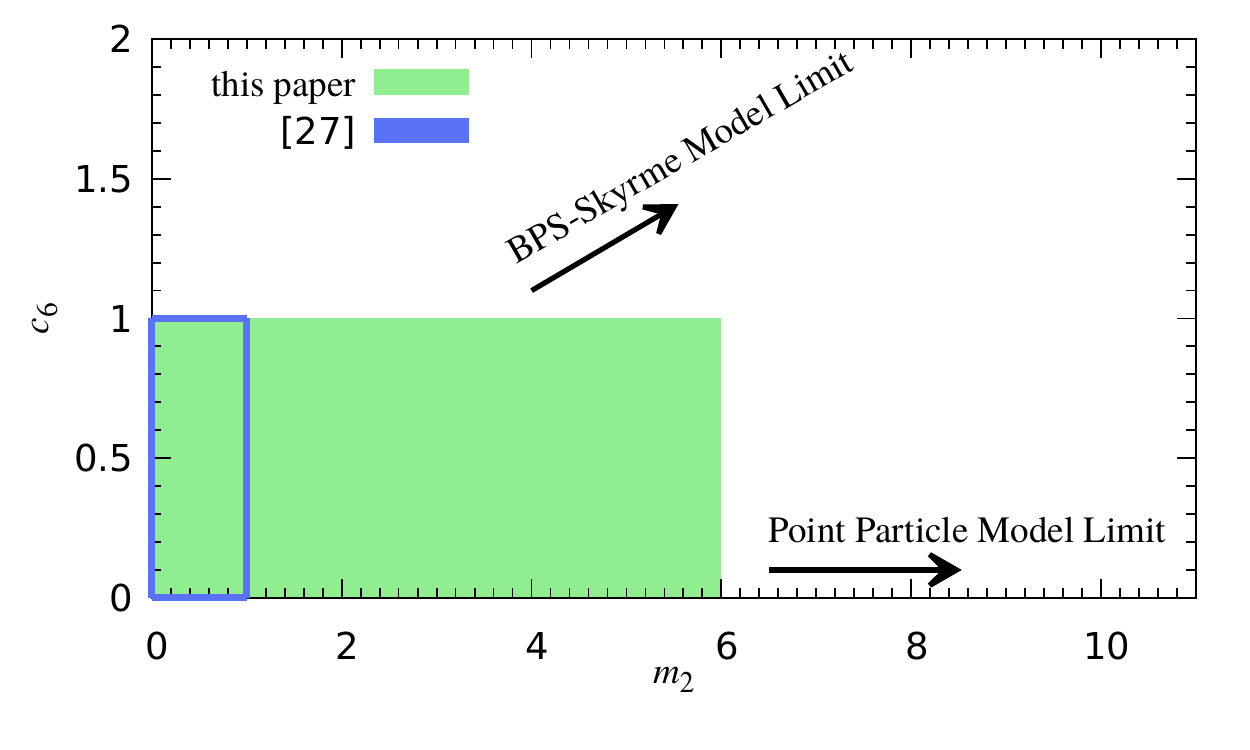}
\caption{Parameter space of the model explored in this paper (green)
as well as what was studied in ref.~\cite{Gudnason:2016tiz} (region
inside the blue box).
The BPS-Skyrme Model Limit (BSML) is defined as
$c_6\propto m_2\to\infty$ and the Point Particle Model Limit (PPML) is
$c_6=0$, $m_2\to\infty$.
}
\label{fig:parmspace}
\end{center}
\end{figure}

In this paper, we perform the full PDE calculations for the 4-Skyrmion
in the generalized Skyrme model with the loosely bound potential and a
standard pion mass term, for small values of the coefficient of the
BPS-Skyrme term and up to large values of the mass parameter of the
loosely bound potential, see fig.~\ref{fig:parmspace}.
In particular, in this paper we study the following region of
parameter space: $c_6\in[0,1]$ and $m_2\in[0,6]$.
Note that it is $m_2^2$ that enters the Lagrangian and hence $6^2=36$
is much larger than the other coefficients in the Lagrangian (which
are all of order one).
In fact, we are close to the limit of how far we can push $m_2$ with
the current numerical codes.
We have not considered the direction of large $c_6$ in this work, as
it will not reduce the binding energy unless we also turn on a large
coefficient of the potential.  
That situation, however, yields two possibilities: either we do not go
beyond the limit where the numerics becomes difficult as discussed
above, or one has to take the near-BPS limit carefully, which will
require overcoming further technical obstacles than dealt with here.
Nevertheless, in the part of parameter space we have studied in this
paper, we are able to obtain a classical binding energy of the
4-Skyrmion as low as $0.2\%$ --  about a factor of four smaller than
the experimental value for Helium-4.
After taking into account the quantum correction to the mass of the
nucleon due to the spin contribution (as it is a spin-$\tfrac12$ state
in the ground state), the binding energy increases to about $3.6\%$ --
about a factor of $4.5$ too large.
This fact suggests that we cannot leave out further quantum
corrections to the nuclear masses; we have to take into account
quantum corrections due to massive modes \cite{Barnes:1997gv}, also
called vibrational modes, see 
e.g.~refs.~\cite{Halcrow:2015rvz,Halcrow:2016spb,Gudnason:2018aej}.

The question remains how large a total contribution the inclusion of
the quantum corrections due to the massive modes will give.
It has been assumed all along that the quantization of solitons can
consistently be made in a semi-classical fashion, where the quantum
corrections are small compared to the (large) mass scale of the
soliton.
This expectation is based on the assumption that the fluctuation
spectrum is weakly coupled, even though the soliton is inherently a
non-perturbative object.
The simplest example is to consider the mass correction to the kink
in the $\lambda\phi^4$ model in 1+1 dimensions
\beq
\Lag^{\rm kink} = -\frac{1}{2}(\p_\mu\phi)^2
- \frac{\lambda}{4\hbar^3}\left(\phi^2 - \frac{\hbar m^2}{\lambda}\right)^2.
\eeq
We can estimate the kink mass with the following back-of-an-envelope
estimate: we rescale the length scale $x^\mu\to\hbar x^\mu/m$ and
rescale the field $\phi\to\sqrt{\frac{\hbar}{\lambda}}m\phi$; the
Lagrangian density is now dimensionless with an overall dimensionful
prefactor of $\frac{m^4}{\hbar\lambda}$.
Assuming the kink exists, its mass must be an order one number times
$\frac{m^3}{\lambda}$ where the $\hbar/m$ came from integrating over
$x^1$. 
Considering now the quantum correction due to massive modes, one
obtains an order-one number times $\hbar\omega$, where $\omega^2$ is the
curvature of the effective potential created by the kink
solution \cite{Dashen:1974cj}. That is, the eigenvalue of the
fluctuation around the kink is $\omega^2\sim m^2/\hbar^2$ and it
follows in the harmonic approximation that the quantum energy is
$\hbar\omega\sim m$.
To realize this, it suffices to note that the second variation of the
potential with respect to the field is
$-\frac{m^2}{\hbar^2}+3\frac{\lambda}{\hbar^3}\phi_{\rm soliton}^2$
and that the soliton solution is proportional to
$\hbar^{\frac12}\lambda^{-\frac{1}{2}}m$; the resultant effective potential for the
fluctuations is thus independent of $\lambda$. 
In this example the mass dimensions of $m$ and $\lambda$ are 1 and 2,
respectively.
If $\lambda\ll m^2$ then the perturbation series makes sense and
furthermore, the quantum correction is much smaller than the classical
contribution
\beq
M_{\rm classical}+\delta M \propto \frac{m^3}{\lambda}\left(1
+ \mathcal{O}\left(\frac{\lambda}{m^2}\right)\right).
\eeq
The situation is more complicated in the case of the 3-dimensional
Skyrmions than in the case of simple 1-dimensional kinks (which are
integrable).
First of all, it is not known to the best of the author's knowledge,
how weakly coupled the fluctuation spectrum really is. Comparing to
the kinks, the question would be how small $\lambda^{\rm eff}$ is for
the Skyrmions. 

The scope in this paper, however, will be to focus on reducing the
classical binding energy of the Skyrmions under the constraint of
preserving as much symmetry of the original Skyrmions as possible.
This is thus in the spirit of assuming that the fluctuation spectrum
of the Skyrmions is weakly coupled and thus the classical mass is the
largest contribution to the mass by far; the zero-mode quantization
gives the most important quantum corrections and the remaining modes
give corrections to the mass of the order of magnitude of the
zero-mode contributions or less. 
The reason for preserving as much symmetry as possible is first of all
to be able to keep some of the phenomenological successes of the
Skyrme model already obtained; such as the Hoyle state and its
corresponding rotational band having a slope a factor of 2.5 lower
than that of the ground state \cite{Lau:2014baa}.
Taking into account the vibrational spectrum of the 4-Skyrmions
vibrating between a flat square configuration and a tetrahedral
arrangement was crucial in obtaining the right spectrum of Oxygen-16,
having a large energy splitting between states of the same spins and
opposite parity \cite{Halcrow:2016spb}.
Both of these results would fall apart if the 4-Skyrmion loses its
cubic symmetry.
Finally, and perhaps even more importantly, the larger the symmetry,
the better the chances are that the symmetry can eliminate unwanted
degeneracies, e.g.~the parity doubling found in the $B=5$ cluster
system in the model of ref.~\cite{Gudnason:2018aej}.

The paper is organized as follows.
In the next section we will introduce the model, define the
observables and finally propose an order parameter for a quantitative
measure of the symmetry change.
In sec.~\ref{sec:numres} we will present the numerical results. 
Finally, sec.~\ref{sec:discussion} concludes the paper with a
discussion of the results and what to do next.

\section{The model}

The model we study in this paper is the generalized Skyrme model --
consisting of a kinetic term, the Skyrme term and the BPS-Skyrme term
-- with a pion mass term and the so-called loosely bound potential.
In physical units we have
\beq
\Lag = \frac{\tilde{f}_\pi^2}{4}\Lag_2
+\frac{1}{e^2}\Lag_4
+\frac{4c_2c_6}{c_4^2e^4\tilde{f}_\pi^2}\Lag_6
-\frac{\tilde{m}_\pi^2\tilde{f}_\pi^2}{4m_1^2}V,
\label{eq:L}
\eeq
where the kinetic term, the Skyrme
term \cite{Skyrme:1962vh,Skyrme:1961vq} and the BPS-Skyrme
term \cite{Adam:2010fg,Adam:2010ds} are given by 
\begin{align}
\Lag_2 &= \frac{1}{4}\Tr(L_\mu L^\mu),\\
\Lag_4 &= \frac{1}{32}\Tr([L_\mu,L_\nu][L^\mu,L^\nu]),\\
\Lag_6 &= \frac{1}{144}\eta_{\mu\mu'}
  (\epsilon^{\mu\nu\rho\sigma}\Tr[L_\nu L_\rho L_\sigma])
  (\epsilon^{\mu'\nu'\rho'\sigma'}\Tr[L_{\nu'} L_{\rho'}
  L_{\sigma'}]),
\end{align}
with the left-invariant current defined as
\beq
L_\mu \equiv U^\dag\p_\mu U,
\eeq
in terms of the chiral Lagrangian or Skyrme field $U$ which is related
to the pions as
\beq
U = \mathbf{1}_2\sigma + i\btau\cdot\bpi,
\eeq
with the nonlinear sigma model constraint $\det U=1$ or
$\sigma^2+\bpi\cdot\bpi=1$ and $\btau$ is a 3-vector of the Pauli
matrices.
The Greek indices $\mu,\nu,\rho,\sigma=0,1,2,3$ denote spacetime
indices and we will use the mostly positive metric signature
throughout the paper. 

We will now switch to (dimensionless) Skyrme units following
ref.~\cite{Gudnason:2016cdo,Gudnason:2016tiz} and denote the
quantities having physical units with a tilde.
In particular, for the energy and length scale we will take
$\tilde{E}=\tilde{\lambda} E$ and $\tilde{x}^i=\tilde{\mu} x^i$,
respectively, where we have the units \cite{Gudnason:2016tiz} 
\beq
\tilde{\lambda} = \frac{\tilde{f}_\pi}{2e\sqrt{c_2 c_4}}, \qquad
\tilde{\mu} = \sqrt{\frac{c_2}{c_4}}\frac{2}{e\tilde{f}_\pi},
\eeq
and finally the pion mass in physical units \cite{Gudnason:2016tiz}
\beq
\tilde{m}_\pi = \frac{\sqrt{c_4}}{2c_2} e \tilde{f}_\pi m_1.
\eeq
Hence in dimensionless units, the Lagrangian \eqref{eq:L} reads
\beq
\Lag = c_2\Lag_2 + c_4\Lag_4 + c_6\Lag - V.
\label{eq:L_Sk_units}
\eeq
For a positive definite energy density, we require $c_2>0$, $c_4>0$
and $c_6\geq 0$.\footnote{As long as the Skyrmion is not emanating
from a black hole horizon \cite{Gudnason:2016kuu,Adam:2016vzf}, we 
could also take $c_6>0$ (for scaling stability due to Derrick's
theorem \cite{Derrick:1964ww}) and $c_4\geq 0$; in this paper,
however, we will fix $c_4>0$. } 

The potential we will consider in this paper is due to the results of
ref.~\cite{Gudnason:2016mms}, which showed that the pion mass term
squared lowers the binding energy further than the pion mass term to
the fourth power while keeping the symmetries of the 4-Skyrmion.
We will also include the standard pion mass term and thus the total
potential is
\beq
V = V_1 + V_2,
\eeq
where we have defined
\beq
V_n \equiv \frac{1}{n}m_n^2(1 - \sigma)^n,
\eeq
and $\sigma=\tfrac{1}{2}\Tr[U]$.
Only $V_1$ gives a contribution to the pion mass.
Both $V_1$ and $V_2$ break explicitly the chiral symmetry
SU(2)$_{\rm L}\times$SU(2)$_{\rm R}$ to the diagonal SU(2)$_{\rm L+R}$.
The target space is thus
SU(2)$_{\rm L}\times$SU(2)$_{\rm R}/$SU(2)$_{\rm L+R}\simeq$
SU(2) $\simeq S^3$.
Since the Skyrmion, which is identified with the baryon, is a
texture \cite{Vilenkin:2000}, it is characterized by the topological
degree, $B$, 
\beq
\pi_3(S^3)=\mathbb{Z}\ni B,
\eeq
where $B$ is called the baryon number.
The baryon number or topological degree of a Skyrmion configuration
can be calculated by
\beq
B = \frac{1}{2\pi^2}\int d^3x\; \mathcal{B}^0, \qquad
\mathcal{B}^{\mu} = -\frac{1}{12}\epsilon^{\mu\nu\rho\sigma}
  \Tr(L_\nu L_\rho L_\sigma).
\eeq
We will throughout the paper denote Skyrmions of degree $B$ as
$B$-Skyrmions. 

Finally, for the numerical calculations we have to settle on a choice
of normalization of the units and we follow that of
refs.~\cite{Gudnason:2016mms,Gudnason:2016cdo,Gudnason:2016tiz},
\beq
c_2 = \frac{1}{4},\qquad
c_4 = 1,
\label{eq:c_coeff}
\eeq
and hence the energies and lengths are given in units of
$\tilde{f}_\pi/e$ and $1/(e\tilde{f}_\pi)$, respectively, while the
physical pion mass is given by 
\beq
\tilde{m}_\pi = \frac{\sqrt{c_4}}{2c_2} e \tilde{f}_\pi
\left.\sqrt{-\frac{\p V}{\p\sigma}}\right|_{\sigma=1}
= 2 e \tilde{f}_\pi m_1,
\eeq
where we have used the coefficients \eqref{eq:c_coeff}. 
In this paper we will use $m_1=1/4$ \cite{Gudnason:2016mms}.

\subsection{Observables}

In this section we will list the observables to be measured in the
numerical calculations. Since they are greatly overlapping with our
previous studies, we will only review them briefly here and refer to
ref.~\cite{Gudnason:2016tiz} for details.
As in refs.~\cite{Gudnason:2016cdo,Gudnason:2016tiz} we will only
consider the 4-Skyrmion in this paper, as it plays a unique role in
the alpha-particle interpretation of the Skyrme model and it is the
building block of the lattice structure appearing for large
nuclei \cite{Battye:2006na,Feist:2012ps,Lau:2014baa,Halcrow:2016spb,Rawlinson:2017rcq}. 
More importantly, it is where to look for the change in symmetry that
inevitably kicks in for strongly repulsive potentials, see e.g.~the
point particle model \cite{Gillard:2015eia,Gillard:2016esy} and also
ref.~\cite{Gudnason:2016mms}. 

As usual we are interested in the classical and spin/isospin
quantum-corrected binding energies. 
The energy of the 1-Skyrmion is obtained by minimizing the static
energy corresponding to (minus) the Lagrangian \eqref{eq:L_Sk_units}
for the hedgehog
\beq
U = \mathbf{1}_2 \cos f(r) + i\hat{\mathbf x}\cdot\btau \sin f(r),
\label{eq:hedgehog}
\eeq
where $\hat{\mathbf x}\equiv {\mathbf x}/r$ is the unit 3-vector at
the origin and $r=\sqrt{\mathbf{x}\cdot\mathbf{x}}$ is the radial
coordinate. 
We will call the energy of the $B$-Skyrmion $E_B$. 
As the initial condition for the 4-Skyrmion, we will use the rational
map Ansatz \cite{Houghton:1997kg}
\begin{align}
U &= \mathbf{1}_2\cos f(r) + i\mathbf{n}_R\cdot\btau \sin f(r),\label{eq:RM1}\\
\mathbf{n}_R &= \left(
  \frac{R+\bar{R}}{1+|R|^2},
  \frac{i(\bar{R}-R)}{1+|R|^2},
  \frac{1-|R|^2}{1+|R|^2}
  \right),\\
R(z) &= \frac{z^4 + 2\sqrt{3}i z^2 + 1}{z^4 - 2\sqrt{3}i z^2 + 1},
\label{eq:RM3}
\end{align}
for the $c_6=0$, $m_2=0$ solution and
$z=e^{i\phi}\tan(\tfrac{\theta}{2})$ is the Riemann sphere coordinate.
Once a numerical solution has been obtained, we can calculate the
classical relative binding energy (CRBE) of the 4-Skyrmion as
\beq
\delta_4 = 1 - \frac{E_4}{4E_1},
\label{eq:delta4}
\eeq
and the quantum-corrected relative binding energy (QRBE) as
\beq
\delta_4^{\rm tot} = 1 - \frac{E_4}{4(E_1 + \epsilon_1)},
\label{eq:delta4tot}
\eeq
where $\epsilon_1$ is the quantum correction due to the isospin
quantization of the 1-Skyrmion. 
The CRBE \eqref{eq:delta4} is independent of the physical units and
thus independent of the calibration of the model.
The QRBE \eqref{eq:delta4tot}, on the other hand, after factoring out
the energy units, still depends on the Skyrme coupling $e$.

Calibrating the Skyrme-like models can be done in many ways and often
ways are invented to minimize the problem of over-binding by
compensating with a better calibration.
In this paper, we will not turn to the calibration for compensating
the over-binding, but try to reduce the binding energy by varying the
parameters of the Lagrangian \eqref{eq:L_Sk_units}.
Thus, we will stick with a simple calibration where we set the mass
and size of the 4-Skyrmion to those of Helium-4.
In order to calculate the electric charge radius of the 4-Skyrmion, we
note that the ground state of Helium-4 is an isospin 0 state and thus
the charge radius in the Skyrme model is given entirely by the baryon
charge radius \cite{Gudnason:2016tiz}
\beq
r_4^2 = r_{4,E}^2 = r_{4,B}^2
= \frac{1}{8\pi^2}\int d^3x\; r^2 \mathcal{B}^0.
\eeq
The calibration now reads \cite{Gudnason:2016tiz}
\begin{align}
\tilde{f}_\pi = 2\sqrt{c_2}\sqrt{\frac{r_4 \tilde{M}_{{}^4{\rm
He}}}{\tilde{r}_{{}^4{\rm He}} E_4}}
= \sqrt{\frac{r_4 \tilde{M}_{{}^4{\rm He}}}{\tilde{r}_{{}^4{\rm He}} E_4}},\qquad
e = \frac{1}{\sqrt{c_4}}\sqrt{\frac{r_4 E_4}{\tilde{r}_{{}^4{\rm He}} \tilde{M}_{{}^4{\rm He}}}}
= \sqrt{\frac{r_4 E_4}{\tilde{r}_{{}^4{\rm He}} \tilde{M}_{{}^4{\rm He}}}},
\end{align}
where in the latter expressions, we plugged in the
normalization \eqref{eq:c_coeff} and the experimental data used here
are $\tilde{M}_{{}^4{\rm He}}=3727\;{\rm MeV}$ and
$\tilde{r}_{{}^4{\rm He}}=8.492\times 10^{-3}\;{\rm MeV}^{-1}$.

With the calibration in place, we can now determine the quantum
correction to the mass of the 1-Skyrmion due to (spin/)isospin
quantization (with the normalization \eqref{eq:c_coeff})
\beq
\tilde{m}_N \equiv \tilde{M}_1
= \frac{\tilde{f}_\pi}{e}\left(E_1 + \epsilon_1\right)
= \tilde{E}_1 + \tilde{\epsilon}_1,\qquad
\epsilon_1 = \frac{e^4}{2\Lambda} J(J+1),
\label{eq:mN_epsilon1}
\eeq
where $\Lambda$ is the diagonal component of the isospin inertia
tensor for the 1-Skyrmion: $U_{ij}=\Lambda\delta^{ij}$ where $U_{ij}$
is given in the next section in eq.~\eqref{eq:Utensor}.
For the hedgehog Ansatz \eqref{eq:hedgehog} the expression for
$\Lambda$ reads
\beq
\Lambda = \frac{8\pi}{3}\int dr\; r^2 \sin^2 f\left(
c_2 + c_4 f_r^2 + \frac{c_4}{r^2}\sin^2f
  +\frac{2c_6\sin^2(f) f_r^2}{r^2}
\right).
\eeq
Finally, for the ground state of the proton, $J=\tfrac{1}{2}$ and thus
$\epsilon_1=\frac{3e^4}{8\Lambda}$.
We will also consider the $\Delta$ resonance as a spin-$\tfrac{3}{2}$
excitation of the 1-Skyrmion \cite{Adkins:1983ya} yielding
\beq
\tilde{m}_\Delta = \frac{\tilde{f}_\pi}{e}\left(E_1 + 5\epsilon_1\right).
\label{eq:mDelta}
\eeq
The $\Delta$ resonance is nevertheless problematic in the Skyrme
model, see the discussion.

As the ground state of Helium-4 is a spin-0, isospin-0 state, there is
no quantum correction to the mass due to zero-mode quantization
(although there are corrections due to massive modes, see the
discussion). 

Finally, we will consider the electric charge radius of the proton and
the axial coupling. The details and tensor expressions are given in
ref.~\cite{Gudnason:2016tiz} and we will just state the final results
here
\begin{align}
r_{1,E}^2 &= \frac{1}{2}r_{1,B}^2
+\frac{\int dr\; r^2\left(c_2r^2\sin^2 f + c_4\sin^2(f)\left(\sin^2 f
  + r^2f_r^2\right) + 2c_6\sin^4(f) f_r^2\right)}{2\int dr\;
  \left(c_2r^2\sin^2 f + c_4\sin^2(f)\left(\sin^2 f + r^2f_r^2\right)
  + 2c_6\sin^4(f) f_r^2\right)},\label{eq:r1E}\\
g_A &= -\frac{4\pi}{3}\int dr\; r\bigg[
  c_2(\sin 2f + r f_r)
  +c_4\left(\frac{\sin^2f\sin2f}{r^2}
    +\frac{2\sin^2(f)f_r}{r}
    +\sin(2f)f_r^2\right) \non
&\phantom{=-\frac{4\pi}{3}\int dr\;r\bigg[}
  +\frac{2c_6\sin^2f}{r^2}\left(\frac{\sin^2(f)f_r}{r}
    + \sin(2f)f_r^2\right)\bigg],
\end{align}
where the baryon charge radius is 
\beq
r_{1,B}^2 = -\frac{2}{\pi}\int dr\; r^2\sin^2(f) f_r,
\eeq
and the axial coupling in physical units is given by
\beq
\tilde{g}_A = \frac{g_A}{c_4 e^2} = \frac{g_A}{e^2},
\eeq
which is obtained by multiplying the dimensionless expression by
$\tilde{\lambda}\tilde{\mu}$ and in the last expression we have used
the normalization \eqref{eq:c_coeff}.

Some geometric observables that we will calculate, are the tensors of
inertia corresponding to the spin and the isospin of the 4-Skyrmion.
For the generalized model \eqref{eq:L} they can be written
as\footnote{Although in the form displayed here the contribution due
to the (sextic) BPS-Skyrme term appears with a relative minus sign,
all terms contribute to the tensors with the same sign (positive for
$U$ and $V$ while $W$ vanishes for the 4-Skyrmion). }
\begin{align}
U_{ij} &= -\frac{1}{2}\int d^3x\; \Big(
  c_2 \Tr(T_i T_j)
  +\frac{c_4}{4}\Tr([L_k,T_i][L_k,T_j]) 
  -\frac{c_6}{8}\Tr(T_i[L_k,L_l])\Tr(T_j[L_k,L_l])
  \Big),\label{eq:Utensor}\\
V_{ij} &= -\frac{1}{2}\int d^3x\;
  \epsilon^{i m n}\epsilon^{j p q} x^m x^p \Big(
  c_2 \Tr(L_n L_q)
  +\frac{c_4}{4}\Tr([L_k,L_n][L_k,L_q]) \non
&\phantom{=-\frac{1}{2}\int d^3x\;\epsilon^{imn}\epsilon^{jpq}x^mx^p\Big(}
  -\frac{c_6}{8}\Tr(L_n[L_k,L_l])\Tr(L_q[L_k,L_l])
  \Big),\\
W_{ij} &= \frac{1}{2}\int d^3x\;
  \epsilon^{j m n} x^m \Big(
  c_2 \Tr(T_i L_n)
  +\frac{c_4}{4}\Tr([L_k,T_i][L_k,L_n]) \non
&\phantom{=-\frac{1}{2}\int d^3x\;\epsilon^{jmn}x^m\Big(}
  -\frac{c_6}{8}\Tr(T_i[L_k,L_l])\Tr(L_n[L_k,L_l])
  \Big),
\end{align}
and they enter the kinetic energy of the
Lagrangian \eqref{eq:L_Sk_units} as
\beq
\mathcal{T} = \frac{1}{2}a_i U_{ij} a_j
- a_i W_{ij} b_j
+ \frac{1}{2}b_i V_{ij} b_j,
\eeq
where the isospin and spin angular momenta, respectively, are defined
as 
\beq
a_i \equiv -i\Tr[\tau_i A^{-1} \dot{A}], \qquad
b_i \equiv i\Tr[\tau_i \dot{B} B^{-1}],
\eeq
and they act on the static Skyrme field, $U_0(x^i)$, as
\beq
U = A U_0(R^{i}_{\phantom{i}j} x^j) A^{-1}, \qquad
R^{i}_{\phantom{i}j} = \Tr[\tau^i B \tau_j B^{-1}],
\eeq
where $B(t)$ ($A(t)$) is an SU(2) matrix that transforms the static 
Skyrmion to the (iso)\-spinning Skyrmion.

\subsection{Symmetry}

One could contemplate how to extract the symmetries from a numerical
Skyrmion configuration. One guess could be to use the tensors of
inertia which encode geometrical information about the soliton, in
particular related to its spinning and isospinning.
Another more brute-force attempt could be to take moments of the
energy; schematically $\int (x^1)^{n_1}(x^2)^{n_2}(x^3)^{n_3} \mathcal{E}$.
This, in principle, could extract further geometrical information from
the numerical Skyrmion.

However, we are interested in a particular symmetry, namely 
octahedral\footnote{In this paper, we have used the term ``cubic''
symmetry referring loosely to the symmetry of the cube, which is
octahedral symmetry. This is because the dual of a cube is a
octahedron and the two latter objects share the symmetry: octahedral
symmetry. Throughout the paper we will use both terms
interchangeably. }
symmetry and would like to know when it is broken to its tetrahedral
subgroup. Therefore we can use the transformations of the octahedral
symmetry which are not symmetry transformations of the tetrahedral
subgroup.
In the following, we place a cube such that the Cartesian axes are
perpendicular to 3 of its faces and the origin is at the center of the
cube. 
The tetrahedral symmetry group contains the following transformations:
the identity, three $C_2$ transformations and eight $C_3$
transformations.
The $C_2$ transformations rotate the cube by $\pi$ around one of the
Cartesian axes and the $C_3$ transformations rotate the cube by
$\pm 2\pi/3$ around an axis in the $(1,1,1)$, $(-1,-1,1)$, $(1,-1,-1)$
or $(-1,1,-1)$ direction.
The octahedral symmetry group includes another six $C_2$
transformations as well as three $C_4$ transformations.
The $C_2$ transformations rotate the cube by $\pi$ in the $(1,1,0)$,
$(-1,1,0)$, $(1,0,1)$, $(1,0,-1)$, $(0,-1,1)$ or $(0,-1,-1)$
direction and the $C_4$ transformations rotate the cube by $\pi/2$
around one of the Cartesian axes.
We should choose a transformation among the latter two, which only
resides in the octahedral symmetry group and is lost when only a
tetrahedral subgroup of the symmetry is preserved.

We choose to construct an order parameter for the octahedral symmetry
as follows. Let us choose one of the $C_4$ symmetry transformations,
say about the $x$-axis. The meaning of the Skyrmion possessing such a
discrete symmetry is of course that after rotating the Skyrmion by
$\pi/2$ about the $x$-axis, we must subsequently perform an
appropriate rotation in isospin space to get back to the original
Skyrmion.
The appropriate isospin rotation to follow the (spatial) $C_{4,x}$
rotation is a rotation by $\pi$ in isospin space about the
$\pi_1$-axis.
The 4-Skyrmion possessing octahedral symmetry will be invariant under
these two subsequent transformations, while the tetrahedrally
symmetric 4-Skyrmion will not.
We can thus construct the order parameter for octahedral symmetry as
follows.
We perform the $C_4$ transformation as well as the $C_2$
transformation in isospin space on the Skyrmion and then subtract off
the original Skyrmion, take a 2-norm of the resulting field and
finally integrate over space. If the symmetry is preserved, this
integral vanishes. 
We thus define
\beq
\sigma^{O_h} \equiv \frac{1}{V}\int d^3x\;
  \Tr\left[\left(\left(e^{i\pi\hat{I}_1}e^{\frac{i\pi}{2}\hat{J}_1}-1\right)U\right)^\dag\left(e^{i\pi\hat{I}_1}e^{\frac{i\pi}{2}\hat{J}_1}-1\right)U\right],
\label{eq:sigmaOh}
\eeq
where we have divided by the volume of the Skyrmion,
$V\equiv\frac{4}{3}\pi r_4^3$, in order to get a dimensionless
result.

With all observables at hand we are now ready to turn to the numerical
calculations.

\section{Numerical results}\label{sec:numres}

The numerical calculations in this paper are all carried out on cubic  
lattices of size $121^3$ with a spatial lattice constant of about
$h_x=0.08$ and the derivatives are approximated by a finite difference
method using a fourth-order stencil.
In previous works we were able to use the relaxation method with a
forward-time algorithm, but that turns out to be too slow for the
generalized Skyrme model when including the BPS-Skyrme term (with
nonvanishing $c_6$).
In this paper, therefore, we used the method of nonlinear conjugate
gradients to find the numerical solutions. 
Although standard implementations of the algorithm work smoothly for
small values of the potential parameter $m_2\lesssim 1$, some
nontrivial tweaking and a sophisticated line search algorithm was
needed for ensuring convergence in the large-$m_2$ part of the
parameter space.
In particular, we found a viable solution based on switching between
the Newton-Raphson algorithm and a line search using a quadratic fit
along the search direction of the conjugate gradients method. 

As a handle on the precision we checked that the numerically
integrated baryon charge is captured by the solution to within
$0.15\%$.
In addition to this, we stopped the algorithm when a local precision
of the equation of motion better than $1.7\times 10^{-6}$ was
obtained. 

\begin{sidewaysfigure}[!ht]
\begin{center}
\includegraphics[width=\linewidth]{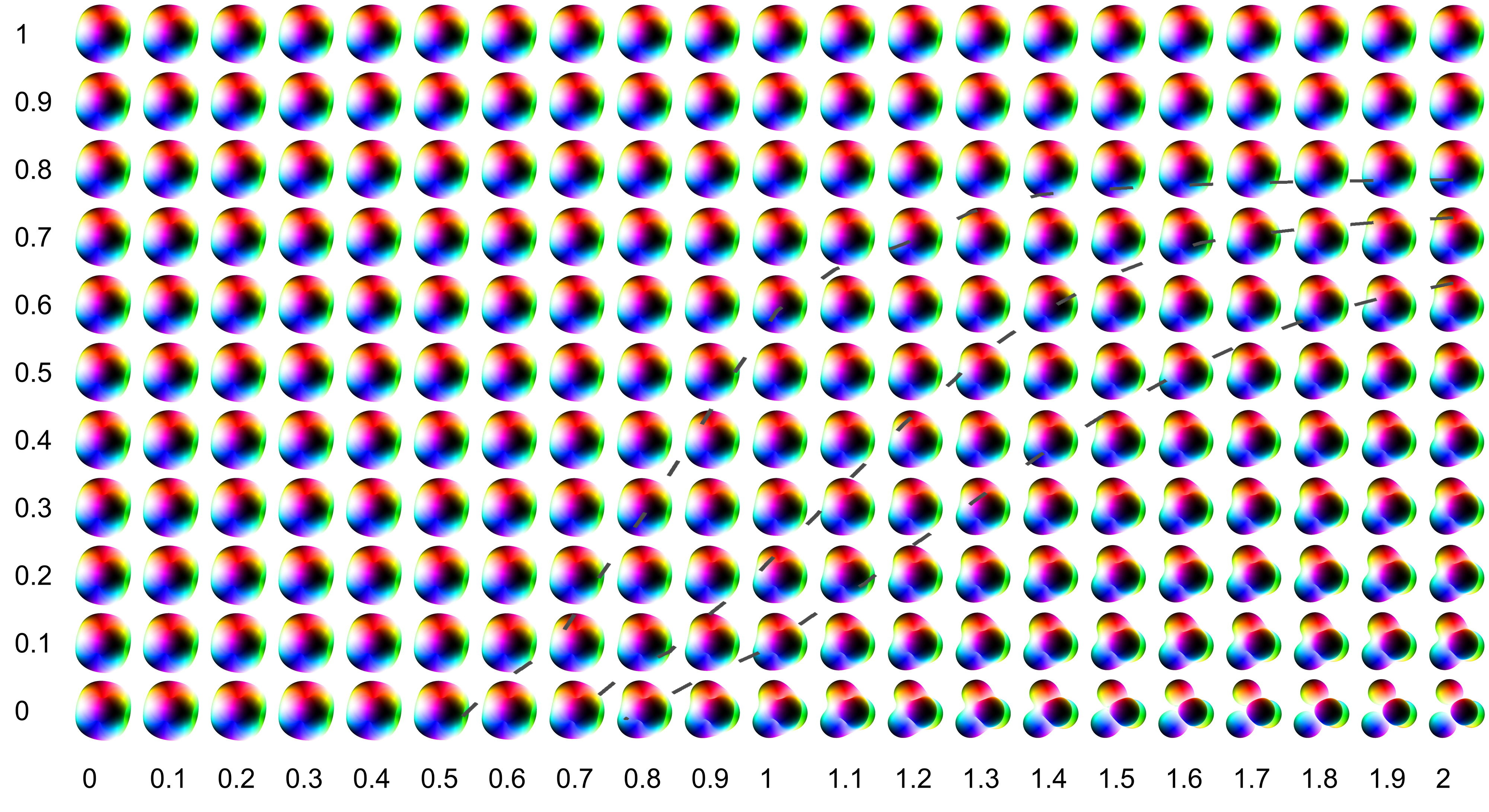}
\caption{Baryon charge density isosurfaces of the 4-Skyrmion solutions
over part of the scanned parameter space. The vertical axis denotes
the values of $c_6$ while the horizontal axis is $m_2$.
The dashed lines show contours of $\sigma^{O_h}=0.1,0.5,1$
from top to bottom.
The center of mass of the Skyrmion corresponds to its position in
parameter space. }
\label{fig:skarr1}
\end{center}
\end{sidewaysfigure}

\begin{sidewaysfigure}[!ht]
\begin{center}
\includegraphics[width=\linewidth]{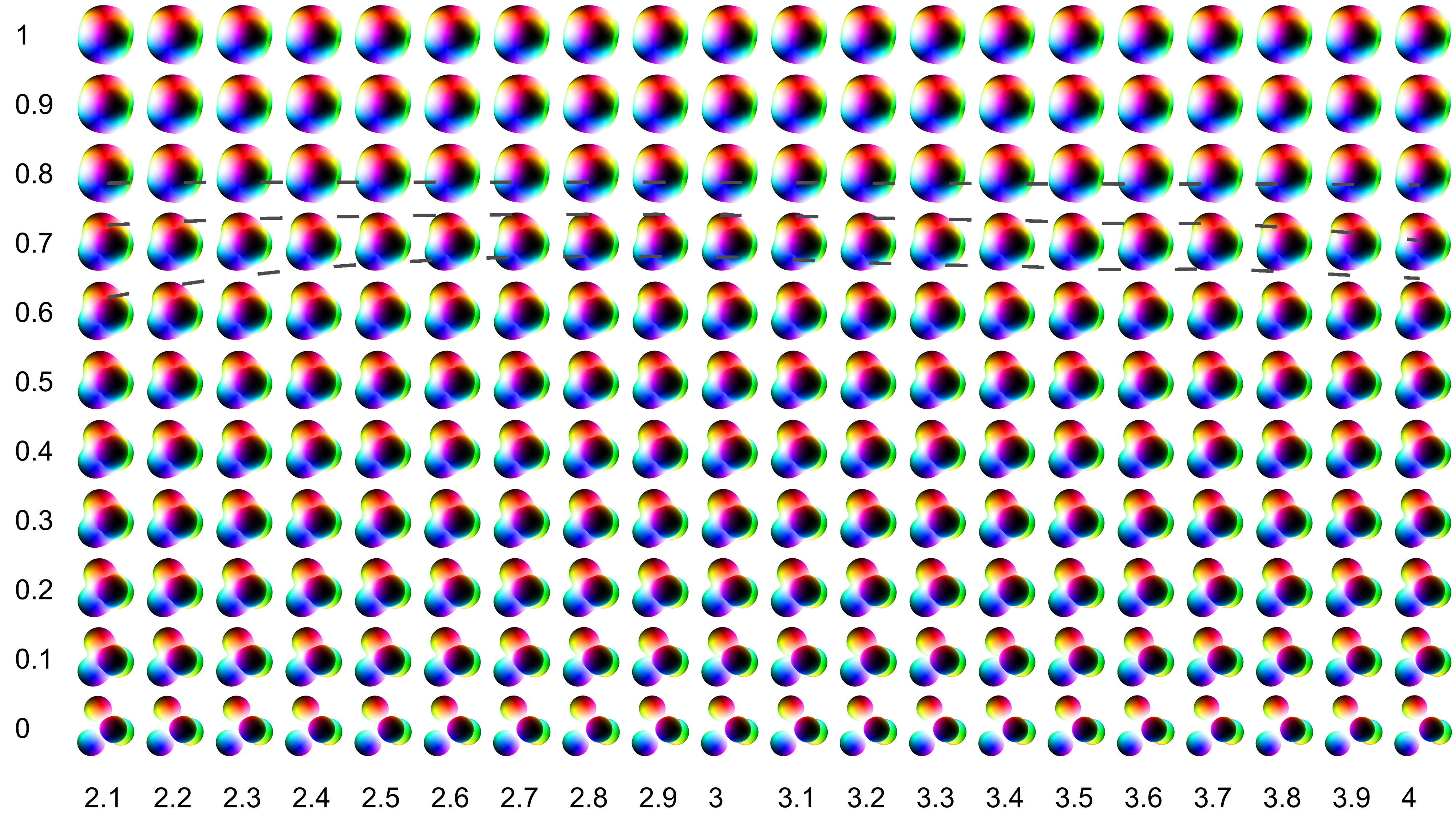}
\caption{Baryon charge density isosurfaces of the 4-Skyrmion solutions
over part of the scanned parameter space. The vertical axis denotes
the values of $c_6$ while the horizontal axis is $m_2$.
The dashed lines show contours of $\sigma^{O_h}=0.1,0.5,1$
from top to bottom.
The center of mass of the Skyrmion corresponds to its position in
parameter space. }
\label{fig:skarr2}
\end{center}
\end{sidewaysfigure}

\begin{sidewaysfigure}[!ht]
\begin{center}
\includegraphics[width=\linewidth]{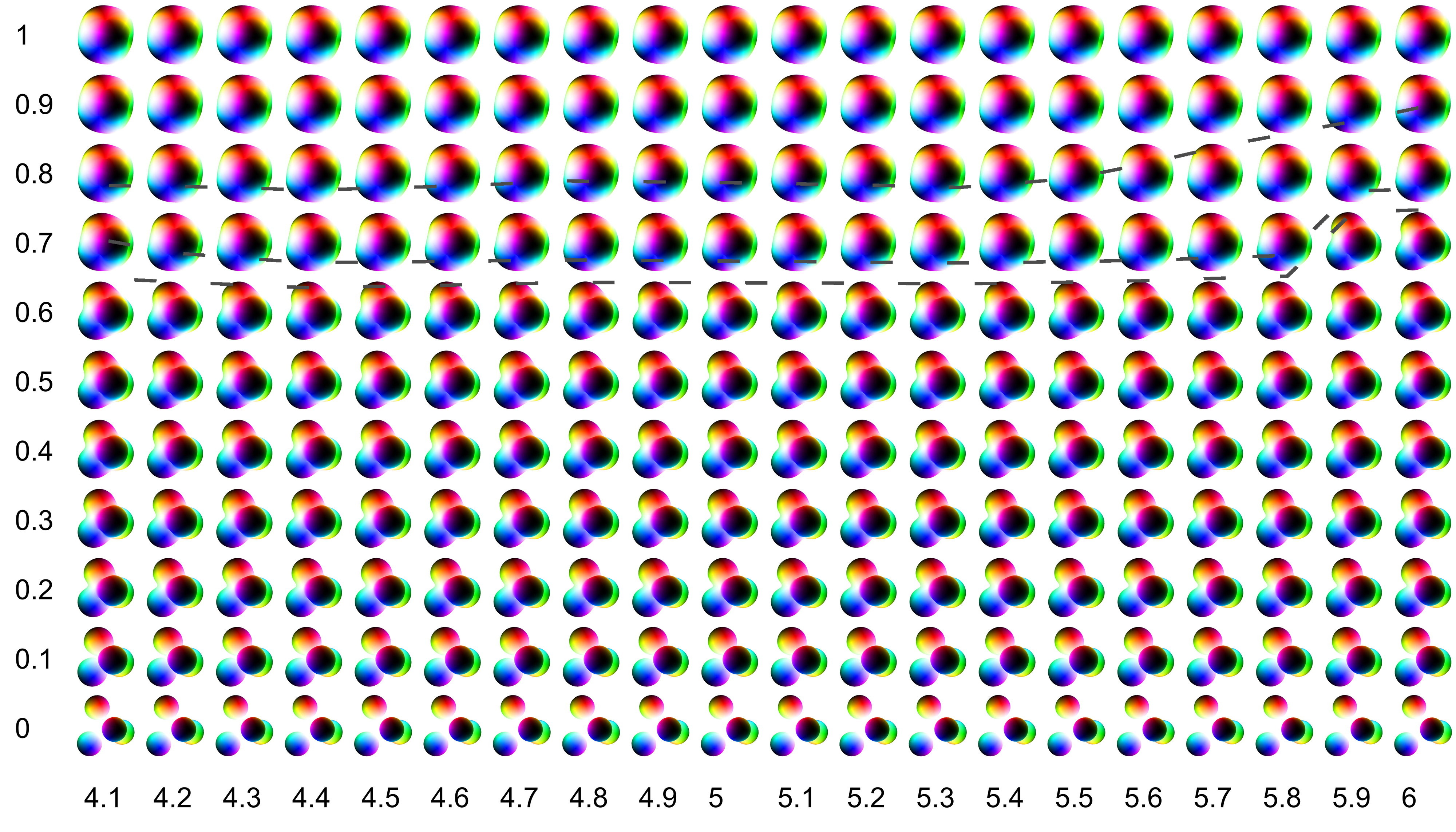}
\caption{Baryon charge density isosurfaces of the 4-Skyrmion solutions
over part of the scanned parameter space. The vertical axis denotes
the values of $c_6$ while the horizontal axis is $m_2$.
The dashed lines show contours of $\sigma^{O_h}=0.1,0.5,1$
from top to bottom.
The center of mass of the Skyrmion corresponds to its position in
parameter space. }
\label{fig:skarr3}
\end{center}
\end{sidewaysfigure}

The solutions obtained and presented here are made on a square grid in
parameter space with $c_6=0,0.1,0.2,\ldots,1$ and
$m_2=0,0.1,0.2,\ldots,6$, yielding a total of 671 numerical solutions. 
The baryon charge density isosurfaces are shown in
figs.~\ref{fig:skarr1}--\ref{fig:skarr3}.
As mentioned in the previous section, the initial condition for the
4-Skyrmion at the point $(m_2,c_6)=(0,0)$ is given by the rational
map Ansatz \eqref{eq:RM1}--\eqref{eq:RM3}. 

The figures in this section are contour plots in the parameter space
with $m_2$ being the abscissa and $c_6$ being the ordinate. 
On all figures, we will overlay three lines with
the order parameter, $\sigma^{O_h}$, which measures if the solutions
possess octahedral symmetry ($\sigma^{O_h}=0$) or remain only
tetrahedrally symmetric ($\sigma^{O_h}>0$). From the top of the
figures and down, the curves correspond to $\sigma^{O_h}=0.1,0.5,1$.

\begin{figure}[!thp]
\begin{center}
\mbox{\subfloat[]{\includegraphics[width=0.49\linewidth]{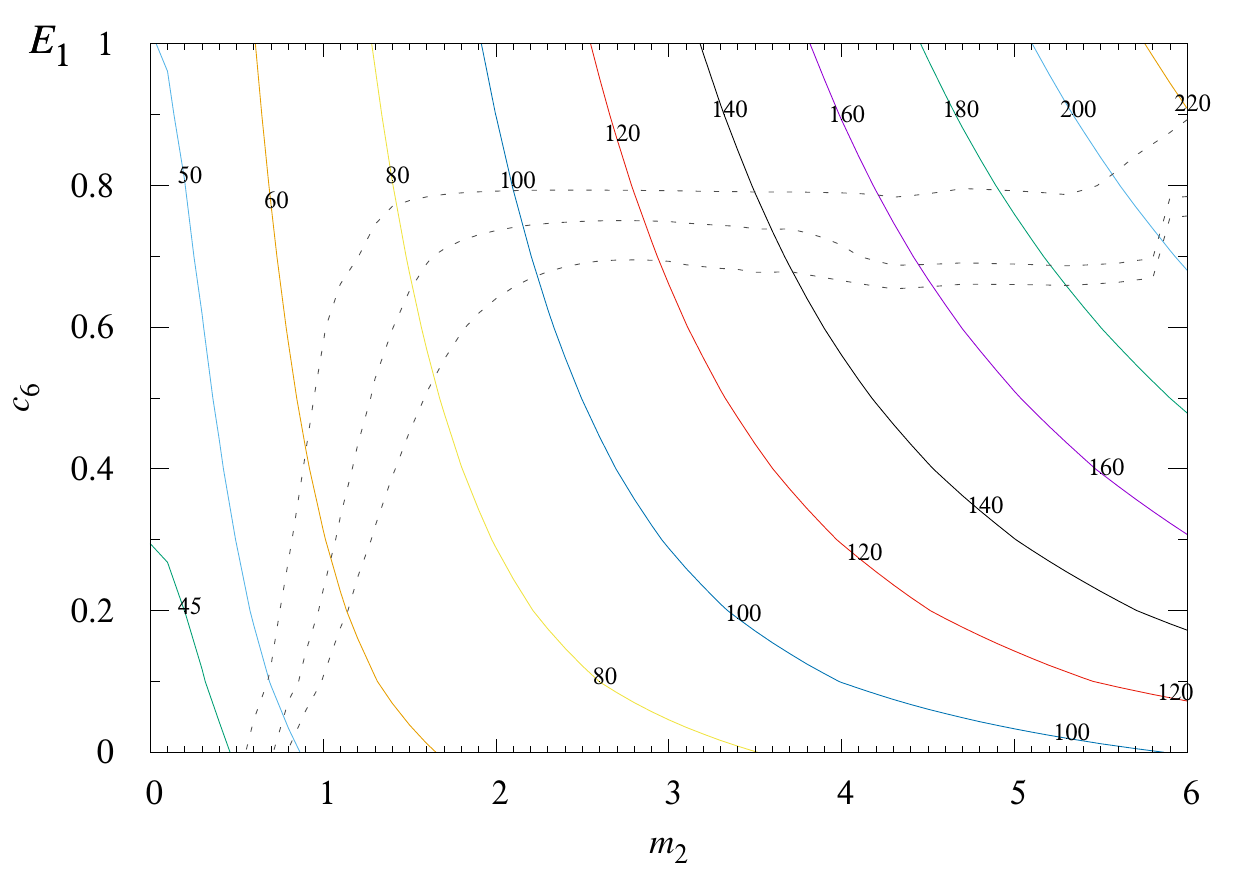}}
\subfloat[]{\includegraphics[width=0.49\linewidth]{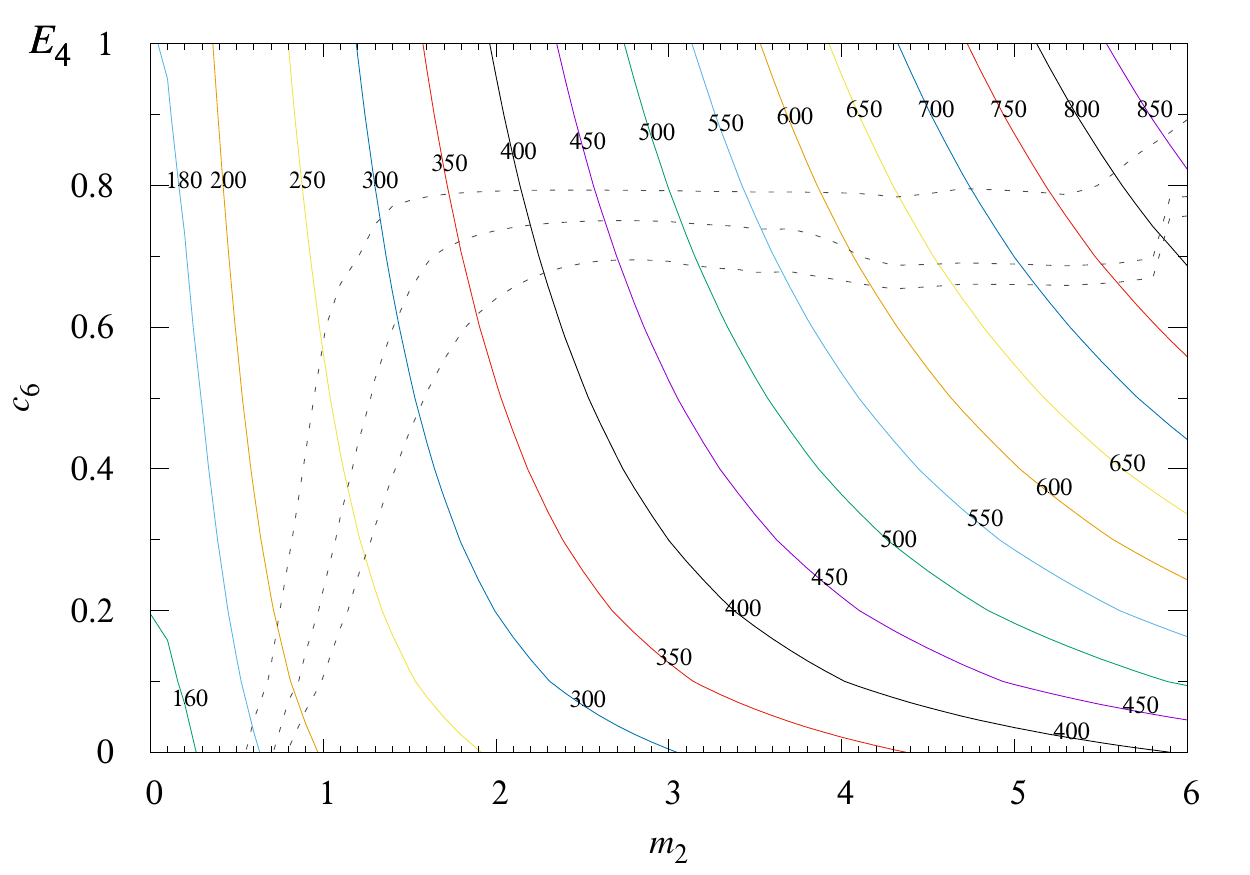}}}
\caption{Energies of the (a) 1-Skyrmion and (b) 4-Skyrmion in Skyrme
  units.
  The dashed lines show contours of $\sigma^{O_h}=0.1,0.5,1$
  from top to bottom. } 
\label{fig:E14}
\end{center}
\end{figure}

\begin{figure}[!thp]
\begin{center}
\mbox{\subfloat[]{\includegraphics[width=0.49\linewidth]{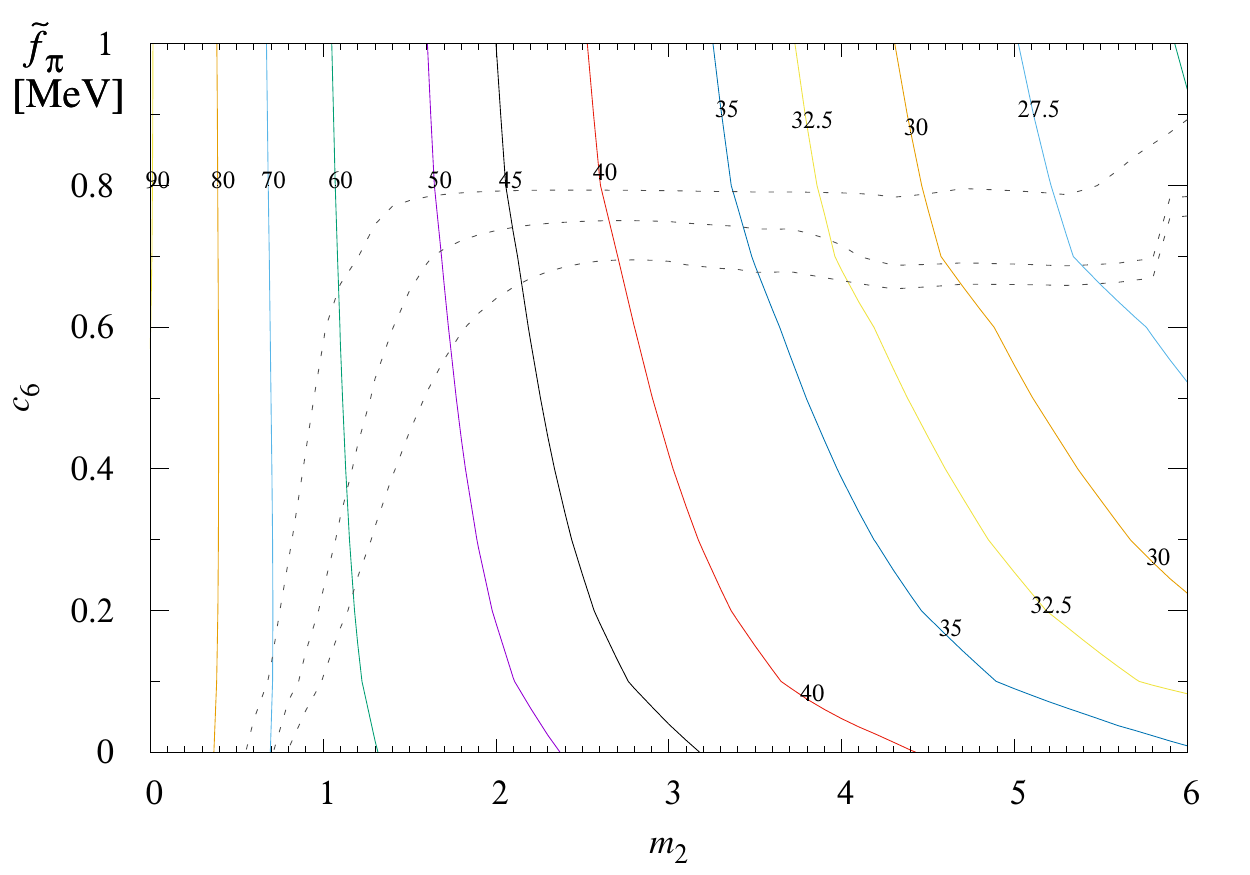}}
\subfloat[]{\includegraphics[width=0.49\linewidth]{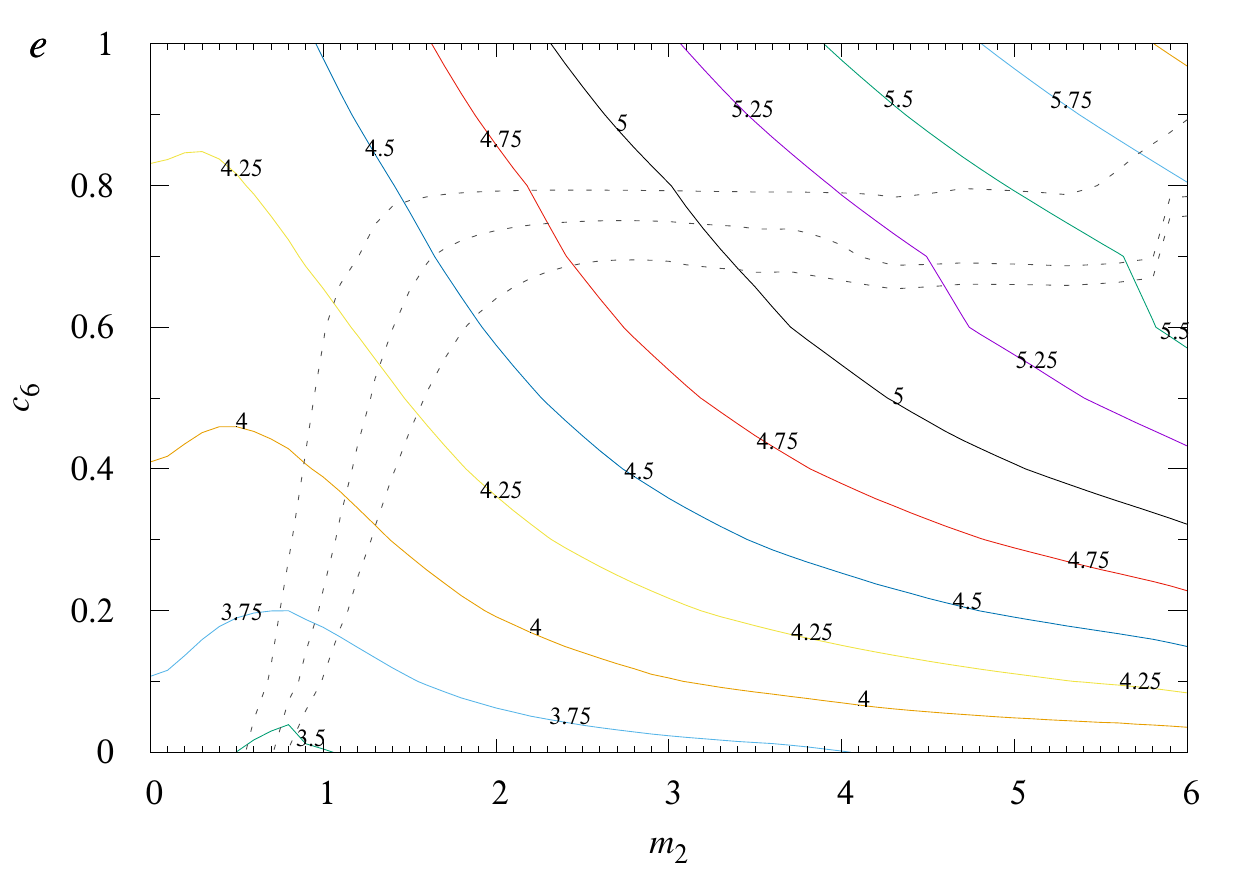}}}
\caption{Calibration constants (a) $\tilde{f}_\pi$ [MeV] and (b) the
  Skyrme coupling constant $e$.
  The dashed lines show contours of $\sigma^{O_h}=0.1,0.5,1$
  from top to bottom. }
\label{fig:fpi_e}
\end{center}
\end{figure}

First we plot the classical energies (masses) of the 1-Skyrmion and
the 4-Skyrmion in Skyrme units in fig.~\ref{fig:E14} just to get a
feel for how the energy changes in parameter space before calibrating
the model.
Both figures show isocurves according to our expectation, i.e.~the
energy increases roughly in quadrature from the contribution due to
the loosely bound potential with coefficient $m_2^2$ and from the
BPS-Skyrme term with coefficient $c_6$.
The increase in energy, nevertheless, is quite drastic.
If we compare the $(m_2,c_6)=(0,0)$ point with the $(m_2,c_6)=(6,1)$
point, the energy increases with a factor of $5.42$ for the 1-Skyrmion
and $5.95$ for the 4-Skyrmion. 

In fig.~\ref{fig:fpi_e} the calibration constants, i.e.~the pion decay
constant, $\tilde{f}_\pi$ and the Skyrme coupling constant $e$, are
shown in the parameter space.
It is interesting that for small $m_2<1$, the pion decay constant
(see fig.~\ref{fig:fpi_e}(a)) with our calibration convention is
almost independent of $c_6$. What happens in this regime is that the
sextic term increases both the size and the energy of the 4-Skyrmion
such that the ratio is almost constant
$r_4(c_6)/E_4(c_6)\propto{\rm const}$. 
For large $m_2$, however, the mass of the 4-Skyrmion increases faster
than the radius and hence the above-mentioned linear relation no
longer holds; as a result the pion decay constant decreases for
increasing $c_6$.
If we now hold $c_6$ fixed, an increase in $m_2$ increases the mass
and reduces the size of the Skyrmions and hence always leads to a
decrease in the pion decay constant.
The combined behavior is displayed in fig.~\ref{fig:fpi_e}(a).
The pion decay constant is underestimated everywhere, since in the
pion vacuum its experimentally measured value is about 184 MeV. 

The Skyrme coupling constant $e$ is shown in fig.~\ref{fig:fpi_e}(b)
and depends on the product of the size and the energy of the
4-Skyrmion and hence displays a different behavior.
For small $c_6\ll 1$, the increase in $m_2$ has a mild behavior since
the loosely bound potential both increases the mass and reduces the
size of the Skyrmions.
For $c_6=0$ and $m_2\lesssim 0.7$ the coupling reduces slightly with
increasing $m_2$, whereas for $m_2>0.7$ the coupling starts to
increase. 
This behavior for constant slices of $c_6$ is continued but the
turning point ($0.7$ above) moves slightly downwards as $c_6$ is
increased. 
For finite $c_6$ the increase in $m_2$ now leads to a larger increase
in the coupling $e$.

\begin{figure}[!tp]
\begin{center}
\mbox{\subfloat[]{\includegraphics[width=0.49\linewidth]{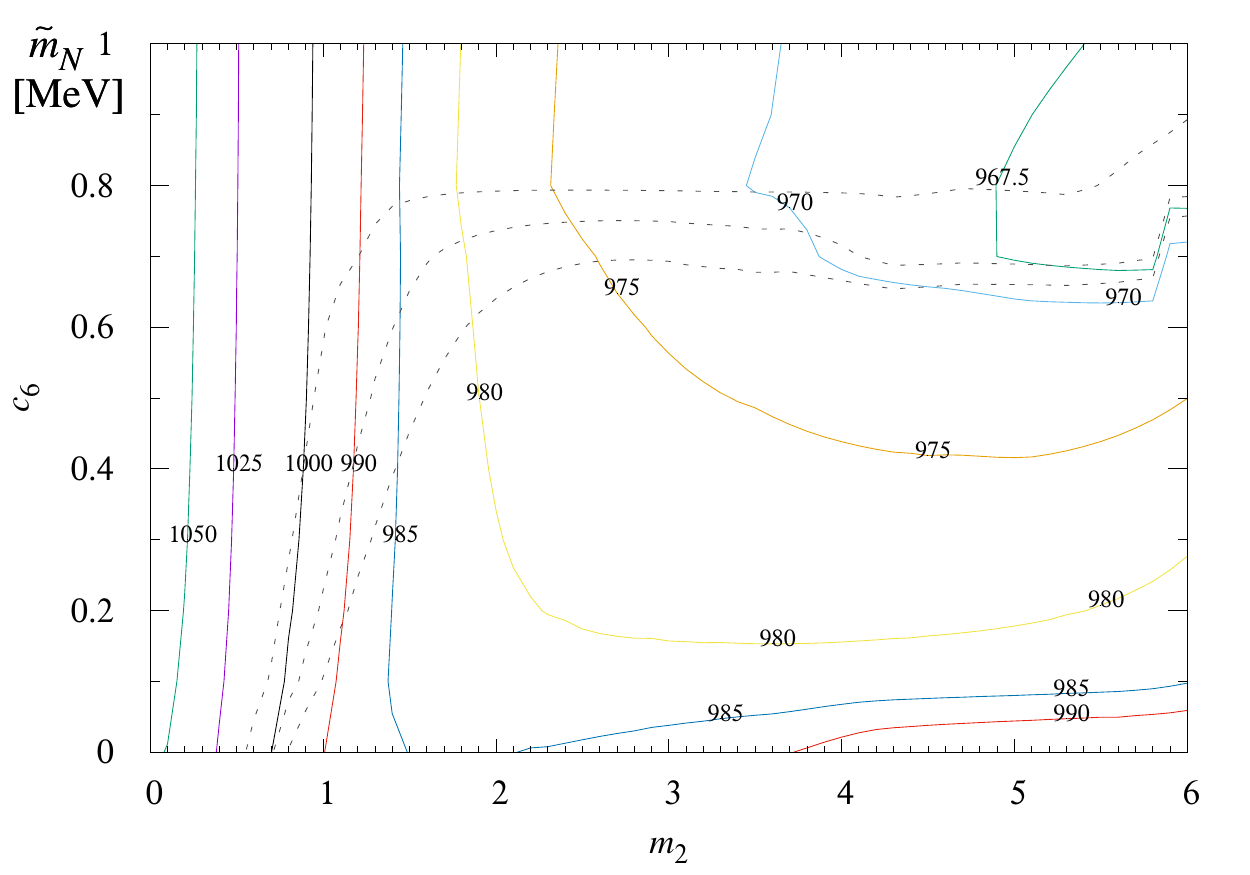}}
\subfloat[]{\includegraphics[width=0.49\linewidth]{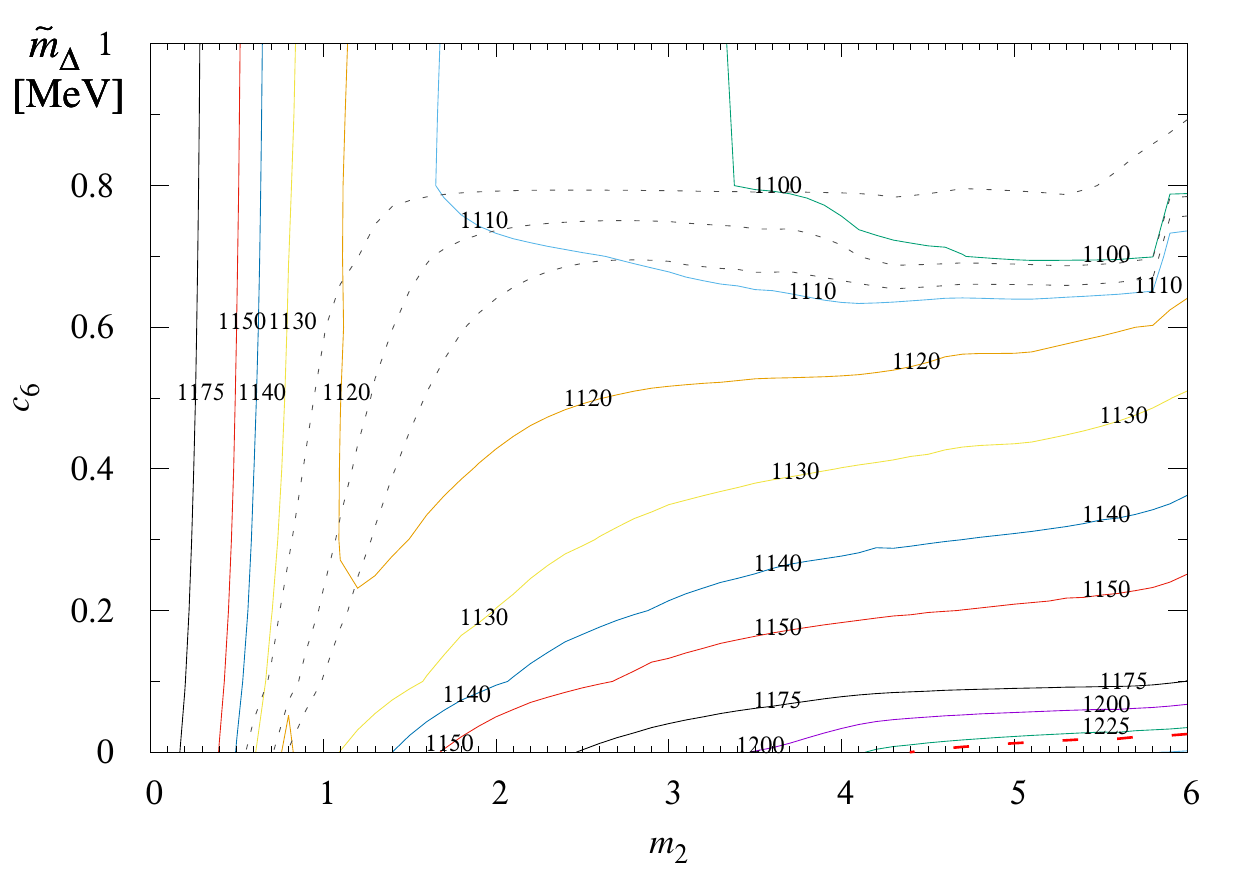}}}
\caption{(a) Nucleon mass, $\tilde{m}_N$ [MeV] and (b) $\Delta$ mass, 
  $\tilde{m}_\Delta$ [MeV].
  The thick red dashed line in (b) is the experimentally measured mass
  of the $\Delta$ resonance ($\tilde{m}_{\Delta}^{\rm exp}\simeq 1232$
  MeV). 
  The dashed lines show contours of $\sigma^{O_h}=0.1,0.5,1$
  from top to bottom. }
\label{fig:mN_mDelta}
\end{center}
\end{figure}

\begin{figure}[!tp]
\begin{center}
\mbox{\subfloat[]{\includegraphics[width=0.49\linewidth]{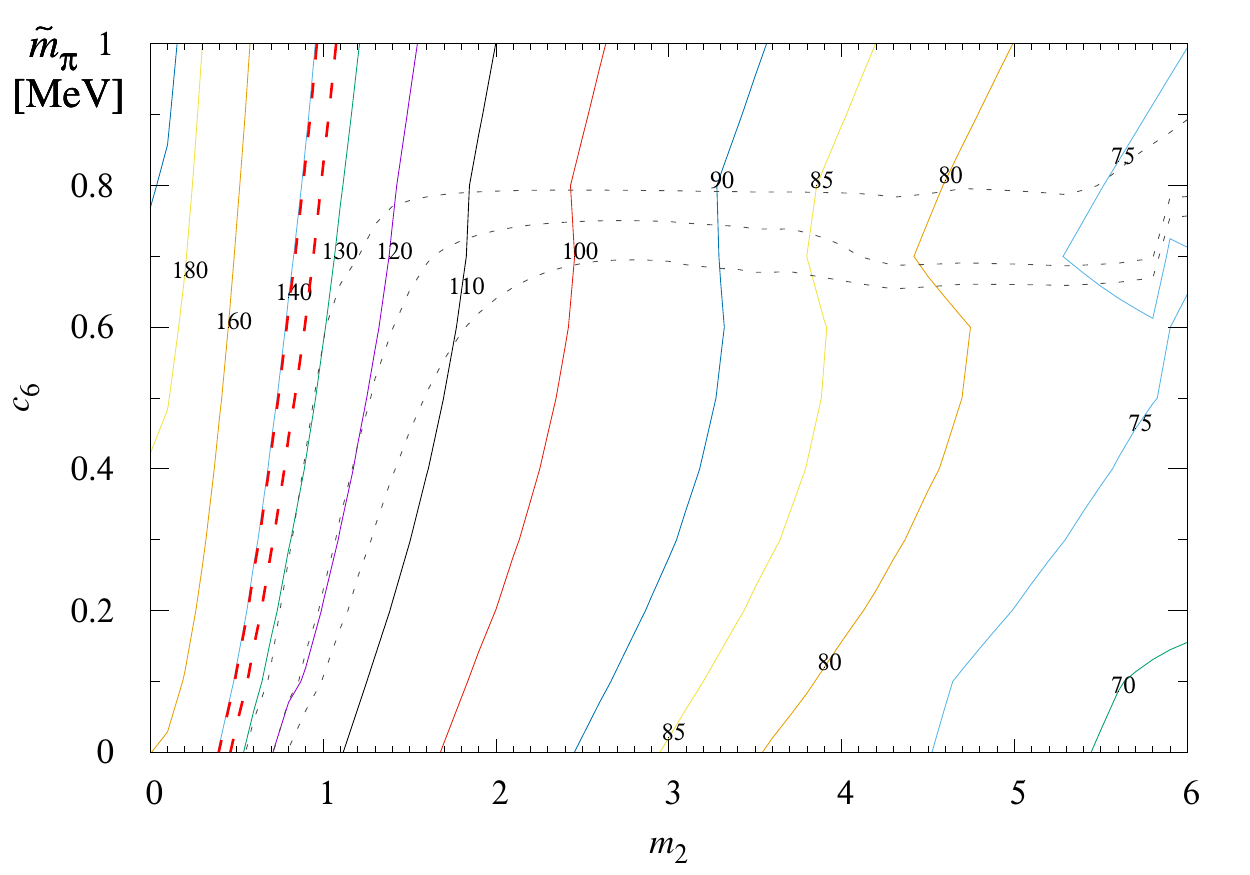}}
\subfloat[]{\includegraphics[width=0.49\linewidth]{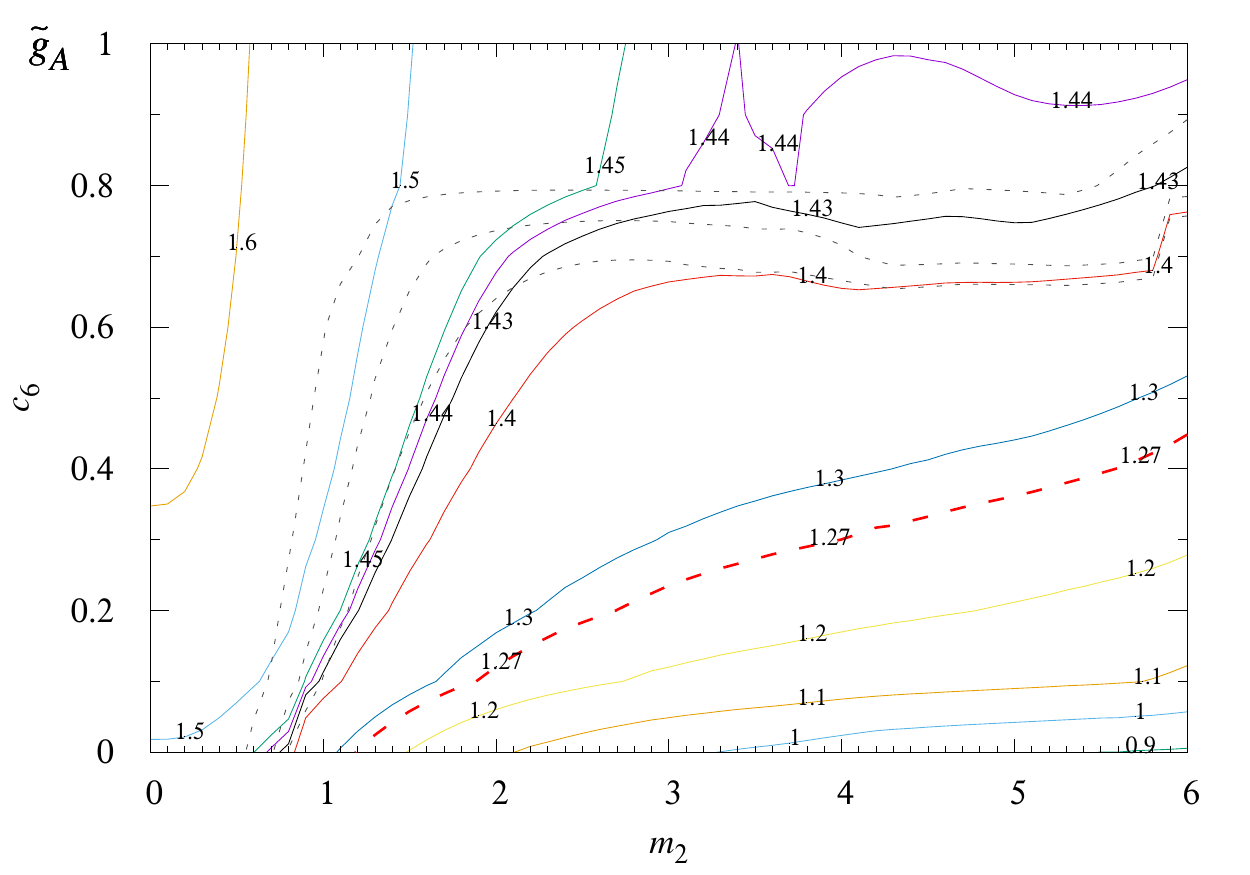}}}
\caption{(a) pion mass, $\tilde{m}_\pi$ [MeV] and (b) axial coupling 
  $\tilde{g}_A$.
  The thick red dashed lines in (a) are the experimentally measured
  pion masses ($\tilde{m}_\pi^{\rm exp}\simeq 135.0\;\textrm{MeV},139.6\;\textrm{MeV}$) and in (b)
  the experimentally measured axial coupling of the nucleon
  ($\tilde{g}_A^{\rm exp}\simeq 1.27$) \cite{PDG:2015}.
  The dashed lines show contours of $\sigma^{O_h}=0.1,0.5,1$
  from top to bottom. } 
\label{fig:mpi_gA}
\end{center}
\end{figure}

We will now show the spectrum of the model, starting in
fig.~\ref{fig:mN_mDelta} with the nucleon mass and the $\Delta$ mass.
The nucleon mass (fig.~\ref{fig:mN_mDelta}(a)) in our calibration
scheme is tightly related to the total binding energy (QRBE), because
we fit the mass and size of the 4-Skyrmion to those of Helium-4.
Therefore, once the binding energy is right, then so is the nucleon
mass.
We will thus discuss this in more detail shortly, but let us mention
that in the top-right part of the parameter space, i.e.~for large
$m_2$ and large $c_6$, the nucleon mass is only overestimated by about
28 MeV, which is about $3\%$ above the experimentally measured value. 

The $\Delta$ mass is shown in fig.~\ref{fig:mN_mDelta}(b).
First of all, we should warn the reader about identifying this spin
excitation of the 1-Skyrmion with the $\Delta$ resonance, as it may be
inherently inconsistent (see also the discussion).
Nevertheless, we will show the results for completeness.
As usual in a Skyrme-like model with this interpretation of the
$\Delta$, its mass is underestimated.
What is worse is that where the binding energy and nucleon mass tend
to their experimentally measured values, the $\Delta$ mass tends to be
the smallest and hence farthest from its true value. 

The pion mass is shown in fig.~\ref{fig:mpi_gA}(a).
In our chosen calibration scheme, the dependence on $c_6$ is mild and
it mostly depends on the value of $m_2$; that is, the pion mass
decreases for increasing $m_2$.
For $m_2\simeq 0.5$--$1$, the pion mass fits well with the
experimentally measured value for our choice of $m_1=1/4$.
In order to get a more physical value of the pion mass in the
top-right corner of the parameter space, we could increase the value
of $m_1$ to say about 0.5 in order to compensate the decrease in the
physical value caused by the loosely bound potential.
We have not done this, as this is not a pressing issue for the model
at the moment; the value of the pion decay constant is very far from
its experimental value and the QRBE is not quite at the physical
values either.
One attitude about the low-energy constants (LECs) is that they should
be ``renormalized'' to the in-medium conditions that the inside of the
baryons possess. If this really justifies the pion decay constant to
differ by more than factors of two (four) from its experiment value,
then the same may apply to the pion mass.

The axial coupling of the nucleon is shown in fig.~\ref{fig:mpi_gA}(b).
In the region of parameter space where the sextic term dominates
($c_6\sim 1$ and $m_2=0$), the axial coupling is generally too large
(however, this may change for larger values of $c_6$ than studied
here), whereas for large $m_2$, $c_6=0$ the axial coupling is
generally too small.
In the top-right corner of parameter space where the binding energy turns
out to be smallest, the axial coupling takes on intermediate values.
Unfortunately, the experimentally measured value (thick red dashed
line) is reached just after the cubic symmetry of the 4-Skyrmion is
lost.

\begin{figure}[!thp]
\begin{center}
\mbox{\subfloat[]{\includegraphics[width=0.49\linewidth]{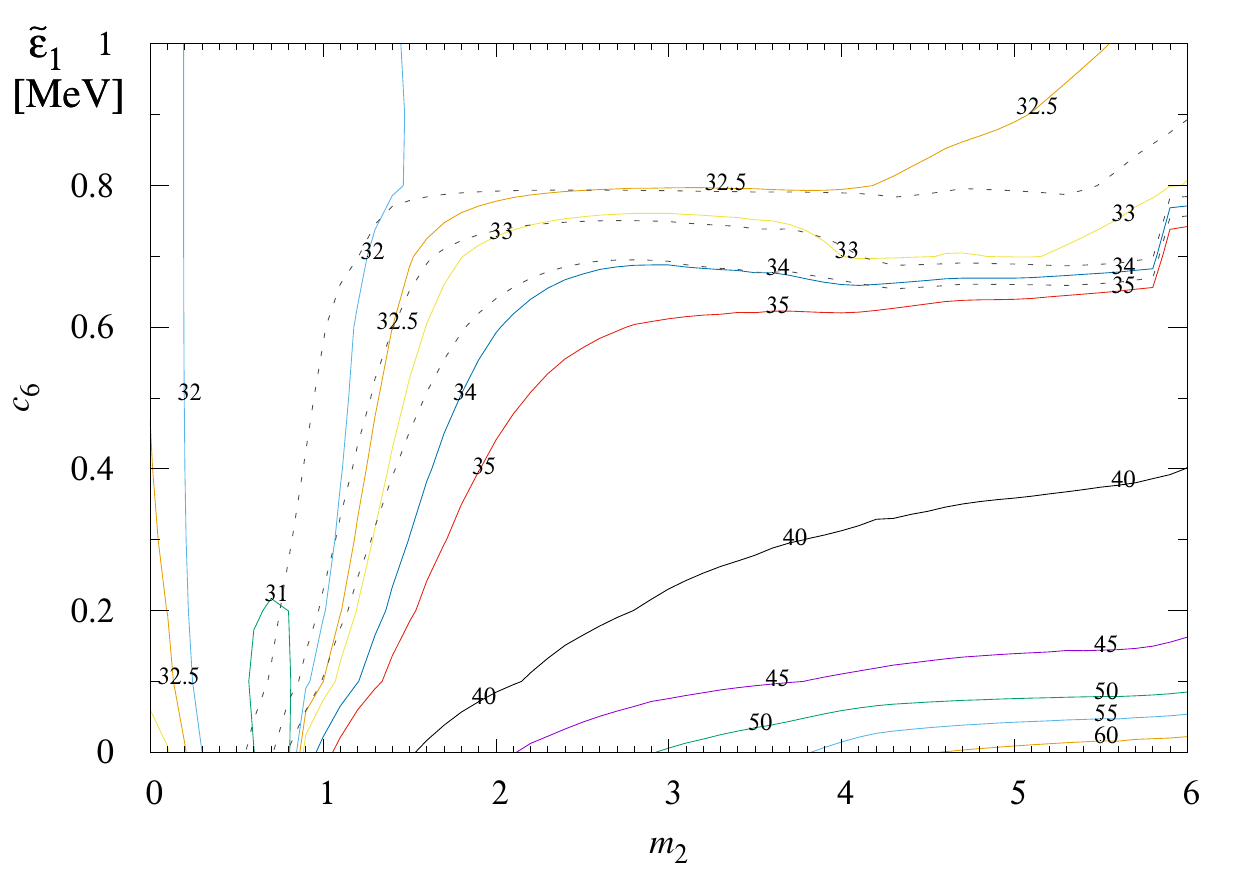}}
\subfloat[]{\includegraphics[width=0.49\linewidth]{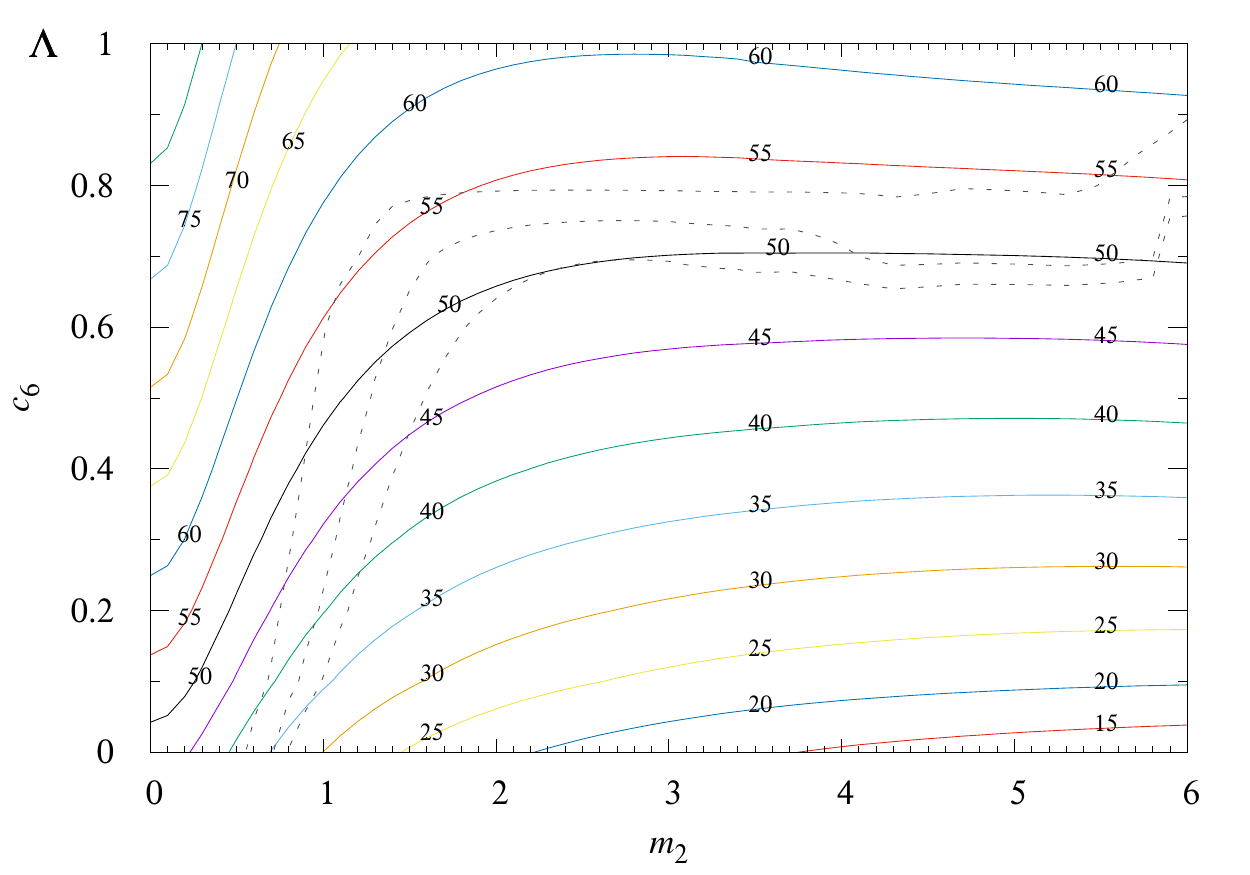}}}
\caption{(a) Quantum isospin correction to the nucleon mass,
  $\tilde{\epsilon}_1$ [MeV] and (b) the diagonal of the isospin
  inertia tensor of the 1-Skyrmion, $\Lambda$, in Skyrme units. 
  The dashed lines show contours of $\sigma^{O_h}=0.1,0.5,1$
  from top to bottom. } 
\label{fig:epsilon1_Lambda}
\end{center}
\end{figure}

\begin{figure}[!thp]
\begin{center}
\mbox{\subfloat[]{\includegraphics[width=0.49\linewidth]{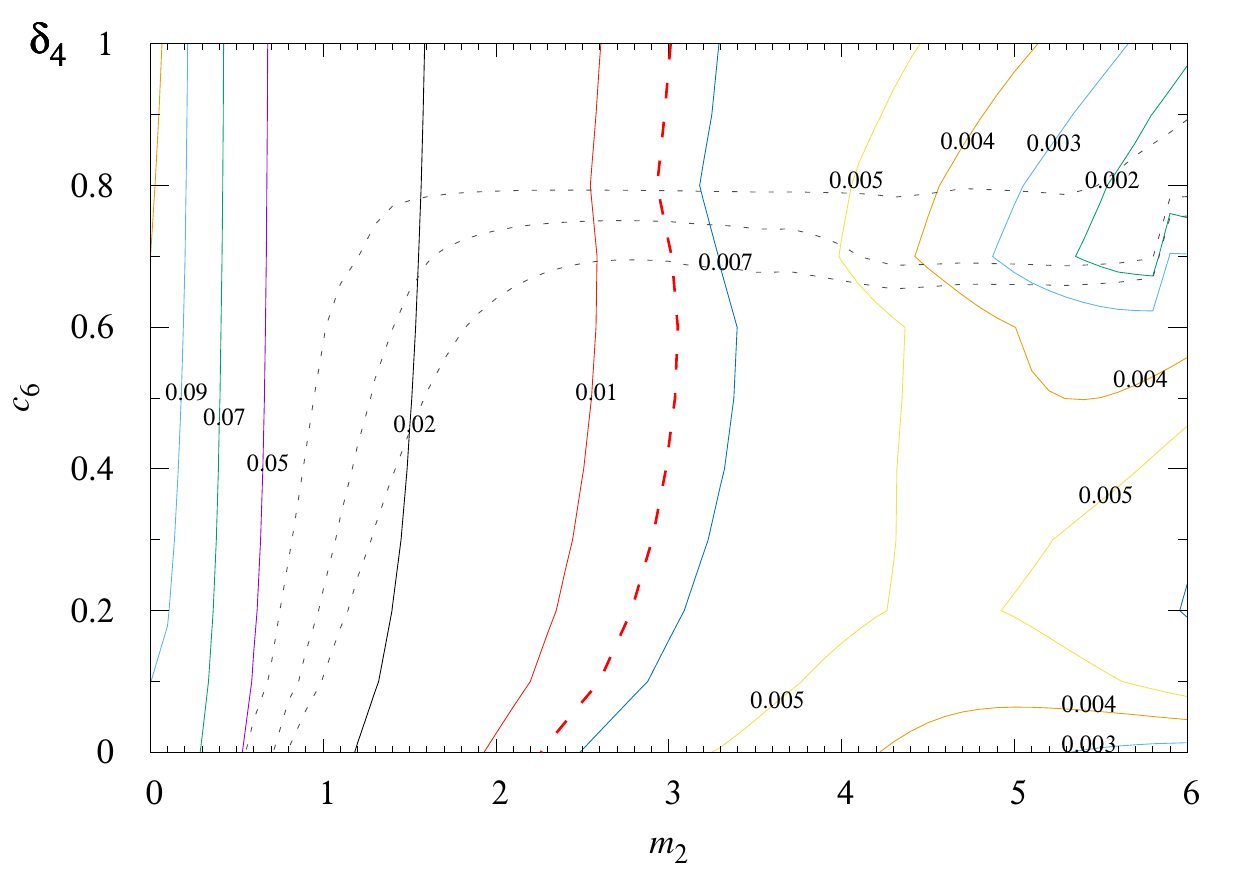}}
\subfloat[]{\includegraphics[width=0.49\linewidth]{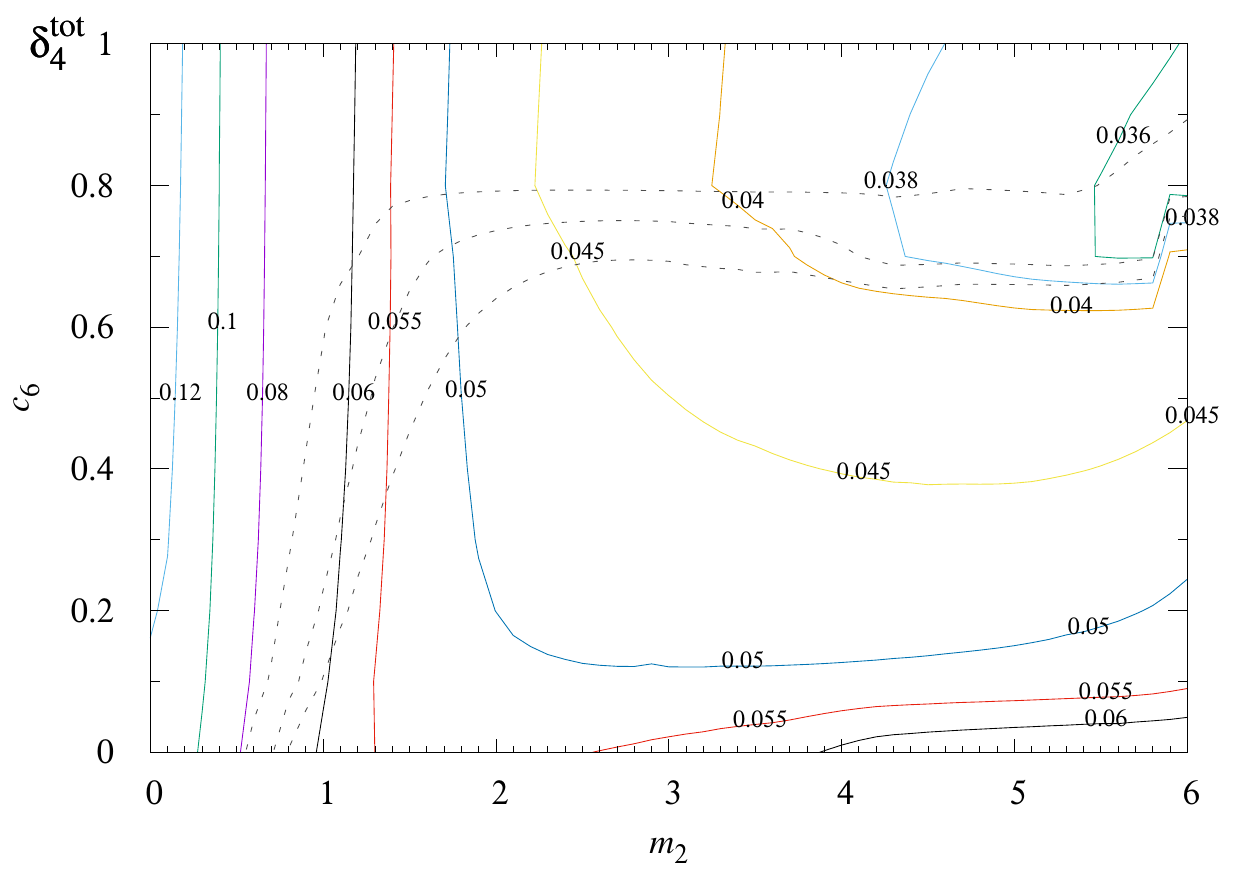}}}
\caption{(a) (CRBE) Classical and (b) (QRBE) quantum relative binding
  energy of the 4-Skyrmion.
  The thick red dashed line in (a) is the experimental binding energy
  of Helium-4 ($\delta_4^{\rm exp}\simeq 0.008$) (it should not be
  compared to the CRBE but to the QRBE; it is shown just for
  reference). 
  The dashed lines show contours of $\sigma^{O_h}=0.1,0.5,1$
  from top to bottom. } 
\label{fig:rbe}
\end{center}
\end{figure}

The quantum mass correction to the nucleon as a 1-Skyrmion is shown in
fig.~\ref{fig:epsilon1_Lambda}(a) and the diagonal value of the
isospin inertia tensor, $\Lambda$ is shown in
fig.~\ref{fig:epsilon1_Lambda}(b).
The two quantities are related by eq.~\eqref{eq:mN_epsilon1}.
As already mentioned, there are two limits where the classical binding 
energy can become infinitesimally small: the point particle model
limit (PPML) which corresponds to $c_6$ fixed, $m_2\to\infty$; and the 
BPS-Skyrme model limit (BSML) which corresponds to $c_6/m_2$ fixed,
$m_2\to\infty$. 
It is interesting to see -- within the calibration scheme adopted here
-- that the direction of the BPS-Skyrme model limit reaches quantum
mass corrections to the 1-Skyrmion which are almost half the values
obtained in the point particle model limit.
If this strategy for obtaining a physically sensible Skyrme model is
correct, it is important to see where the quantum correction to the
1-Skyrmion can be suppressed enough to reach physically measured
values of the binding energies.
This viewpoint is typical for a purist particle physicist who prefers
as little fine tuning as possible.
Other possibilities are of course that the physics at the atomic scale
is highly fine tuned and there are big corrections to both the
1-Skyrmions and the $B$-Skyrmions canceling each other out almost
precisely, leaving behind binding energies at the 1-percent level (see
also the discussion).

The quantity $\tilde{\epsilon}_1$ contains both the energy unit, the
value of the Skyrme coupling and the isospin inertia tensor's diagonal
value.
In order to disentangle the various effects, we show the value of the
diagonal of the isospin inertia tensor, $\Lambda$, in
fig.~\ref{fig:epsilon1_Lambda}(b).
As $\tilde{\epsilon}_1\propto\Lambda^{-1}$, we can see that, overall,
the above mentioned behavior indeed stems from $\Lambda$ and not 
peculiarities of the calibration.
That is, the smallness of the quantum correction in the BPS-Skyrme
model limit comes from the fact that $\Lambda$ is bigger than in the
point particle model limit. This fact, in turn, can be traced directly
to the fact that the Skyrmions become larger when a sizable sextic
term (BPS-Skyrme term) is included and they become smaller when a
strong loosely bound potential is turned on. 

We are now ready to present one of the main results of the paper,
namely the relative binding energies in fig.~\ref{fig:rbe}.
In this part of parameter space, the CRBE (fig.~\ref{fig:rbe}(a)) is
almost independent of $c_6$.
That is, only cranking up the sextic term does not reduce the
CRBE but in fact -- in this calibration scheme -- it leads to a slight
increase in the binding energy. The effect is quite mild though.
The explanation is that the sextic term just makes the Skyrmions
larger and heavier; however, after the calibration this effect is
almost swallowed up.
The loosely bound potential, on the other hand, does its job very
well. The CRBE is reduced to the 1-percent level around
$m_2\sim 2$--$2.6$ depending on the value of $c_6$, and it reaches
values slightly below $0.2\%$ for $m_2\sim 6$.
Now the sextic term is crucial for what happens.
If we solely include the loosely bound potential, the model loses the
platonic symmetries of the Skyrmions and the Skyrmions become well
separated point particles.
However, if we turn on the BPS-Skyrme term, the 4-Skyrmion can retain
its cubic symmetry and possess a CRBE below the 1-percent level.
The upper black dashed line shows the symmetry order parameter for
$\sigma^{O_h}=0.1$ and the cubic symmetry can also be observed in
figs.~\ref{fig:skarr1}--\ref{fig:skarr3}.

The problem of the binding energies is not quite solved yet, because
we have to identify the quantum state of the Skyrmion with the nuclear
particle.
This, in particular, means that all binding energies are relatively
increased by the fact that the nucleon receives a quantum correction
for being a spin-$\tfrac{1}{2}$ particle in the ground state.
This means that if it is a consistent treatment not to include any
other quantum corrections (which is probably not the case), then we
must find a point in the parameter space where the quantum mass
correction to the 1-Skyrmion is at the 1-percent level.
In fig.~\ref{fig:rbe}(b) we can see that the discussion of the quantum 
correction $\tilde{\epsilon}_1$ carries directly over to the QRBE. 
In particular, in the direction of the point particle model limit
($c_6=0$, $m_2\to\infty$) the QRBE we reach in the parameter space is
only as low as slightly below $6\%$, whereas in the direction of the
BPS-Skyrme model limit ($c_6\propto m_2\to\infty$) the QRBE reaches
values as low as $3.6\%$, which for Helium-4 should be compared to
about $0.8\%$ (experimental value), i.e.~about $2.8\%$ over-binding. 
This is, however, the level of over-binding also sometimes present in
nuclear models like the ab initio no-core shell model, see
e.g.~Ref.~\cite{Barrett:2013nh}.

\begin{figure}[!thp]
\begin{center}
\mbox{\subfloat[]{\includegraphics[width=0.49\linewidth]{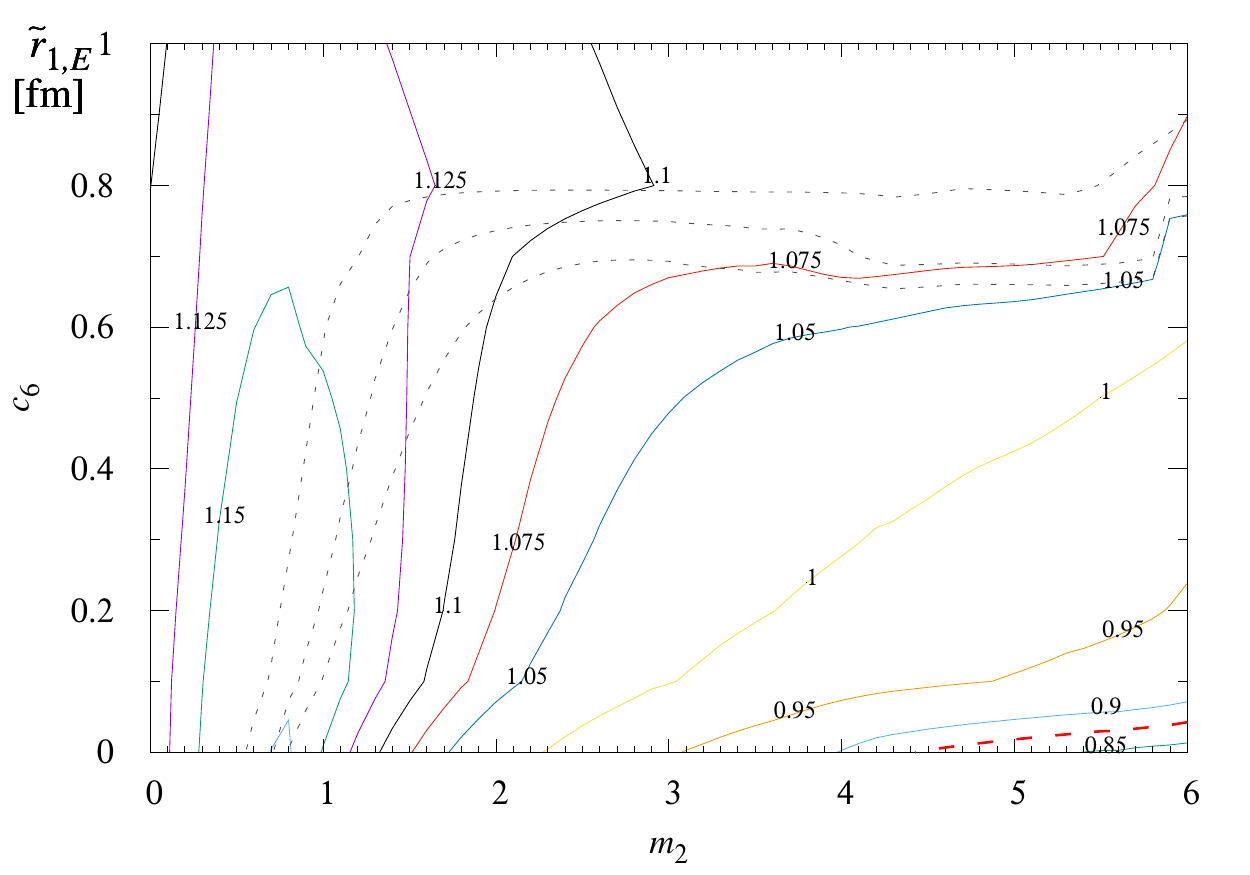}}
\subfloat[]{\includegraphics[width=0.49\linewidth]{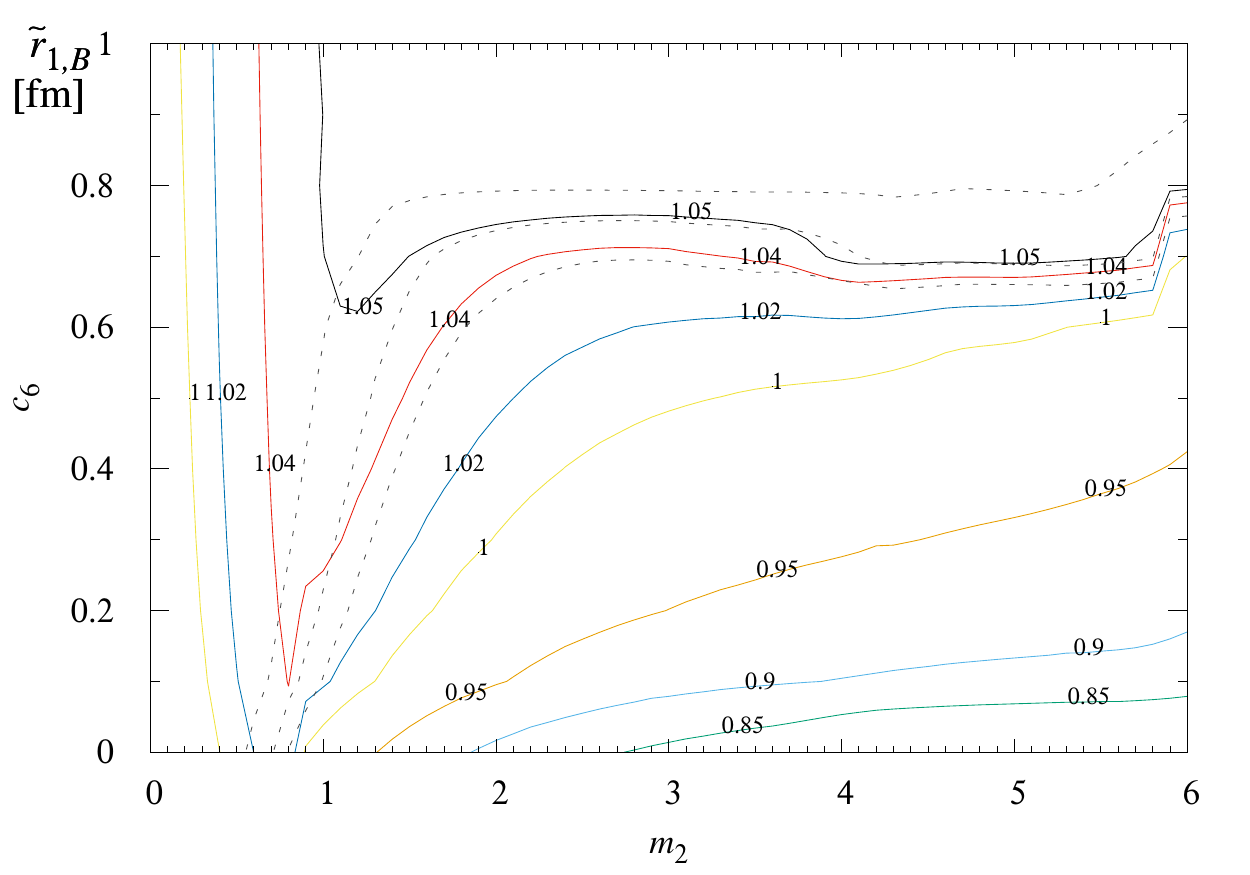}}}
\caption{(a) ($\tilde{r}_{1,E}$) Electric and (b) ($\tilde{r}_{1,B}$)
  baryon charge radii of the proton.
  The thick red dashed line in (a) is the experimentally measured
  charge radius of the proton (CODATA)
  ($\tilde{r}_{1,E}^{\rm exp}\simeq 0.875$ fm) \cite{PDG:2015}. 
  The dashed lines show contours of $\sigma^{O_h}=0.1,0.5,1$
  from top to bottom. } 
\label{fig:r1}
\end{center}
\end{figure}

\begin{figure}[!thp]
\begin{center}
\mbox{\subfloat[]{\includegraphics[width=0.49\linewidth]{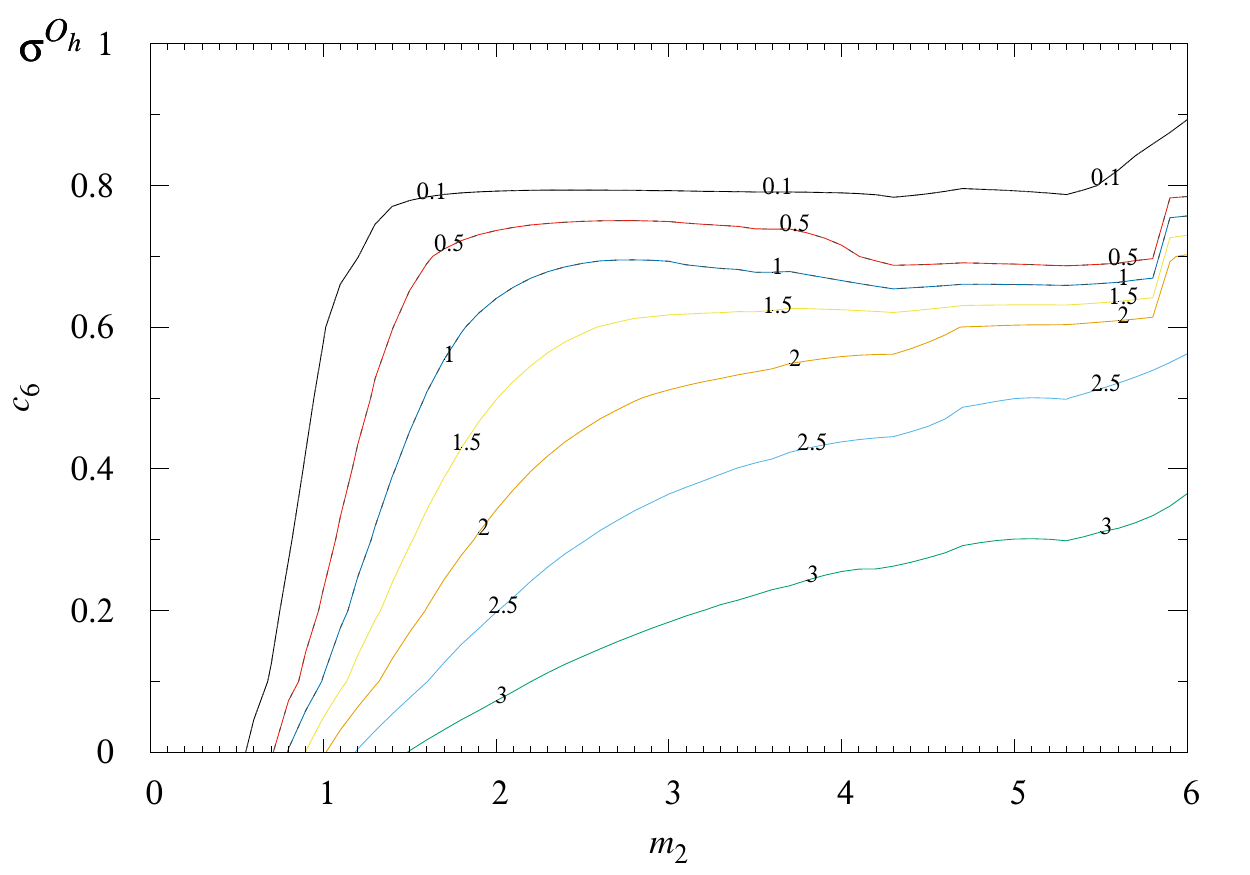}}
\subfloat[]{\includegraphics[width=0.49\linewidth]{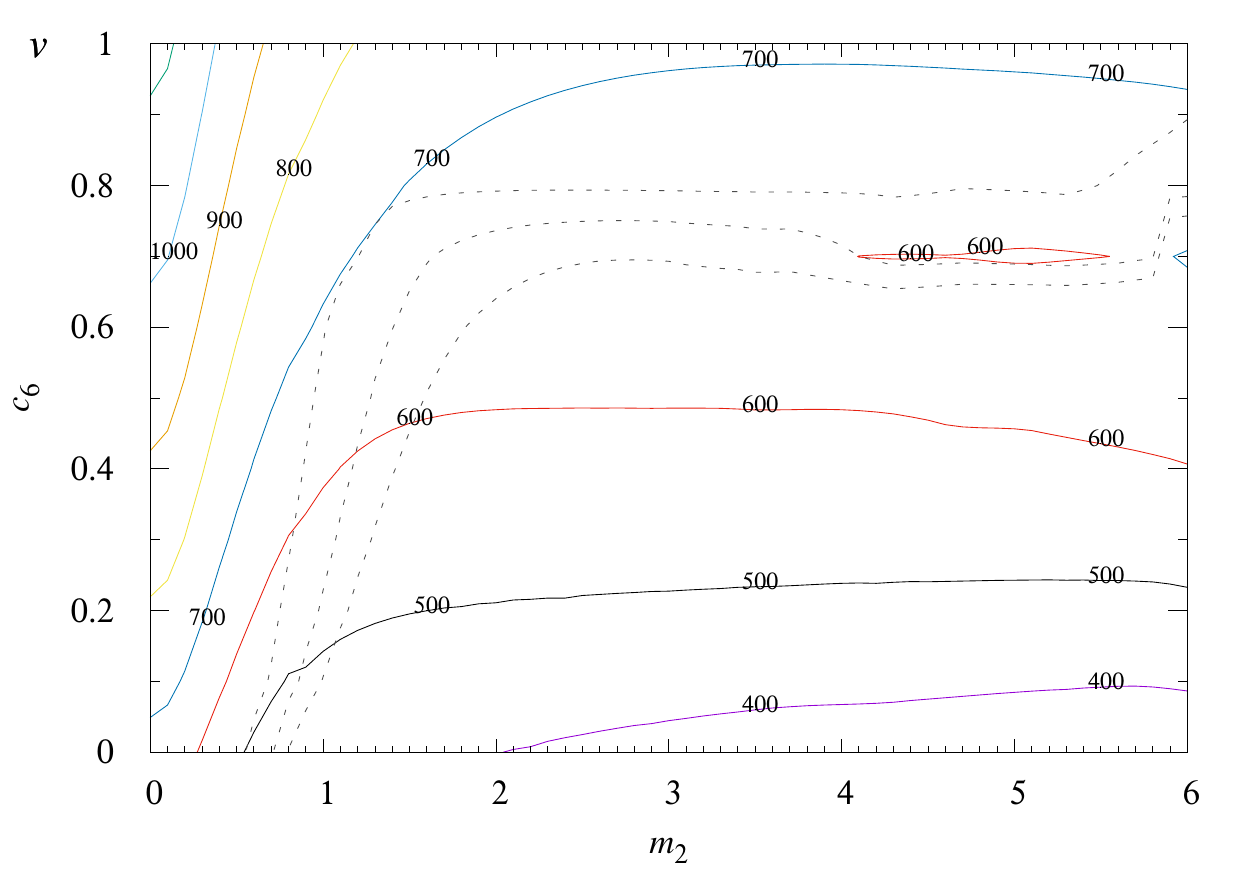}}}
\caption{(a) Order parameter for octahedral symmetry, $\sigma^{O_h}$,
  and (b) diagonal component $v$ of the spin inertia tensor of the
  4-Skyrmion, $V_{ij}=v\delta^{ij}$, in Skyrme units.
  The dashed lines show contours of $\sigma^{O_h}=0.1,0.5,1$ 
  from top to bottom. } 
\label{fig:sigmaOh_v}
\end{center}
\end{figure}

We will now turn to the electric charge radius of the proton, which is
shown in fig.~\ref{fig:r1}(a). In the Skyrme model it has half its
contribution from the baryon charge density and the other half from
the vector charge corresponding to the isospin, see
eq.~\eqref{eq:r1E}.
It is interesting to see that we can obtain the best QRBE for large
$m_2$ and large $c_6$ and hence also the best nucleon mass, but the
physically measured value of the electric charge radius of the proton
is reached spot-on in the direction of the point particle model limit;
that is for $c_6=0$ and $m_2\sim 5$--$6$.
In all other parts of parameter space, the 1-Skyrmion size is
generally overestimated.
This is due to the fact that $B$-Skyrmions tend to be too small and
the 4-Skyrmion is no exception.
Due to the calibration where we fit the size of the 4-Skyrmion to that
of Helium-4, the 1-Skyrmion is thus generally too large.
It is interesting, nevertheless, to see that the point particle model
limit gets the proton size right.

The baryon charge radius is not physically measurable, but it is a
component of the electric charge radius of the proton. For large
$m_2$, we can see that their behaviors are comparable, see
fig.~\ref{fig:r1}(b). 

The last but important observable we will study here is our proposal 
for an order parameter for the symmetry breaking of the cubic
(octahedral) symmetry of the 4-Skyrmion, see eq.~\eqref{eq:sigmaOh}.
By comparing the symmetries observed in the
figures \ref{fig:skarr1}--\ref{fig:skarr3} with the values seen in 
fig.~\ref{fig:sigmaOh_v}(a), we see that for $\sigma^{O_h}<0.1$ the
4-Skyrmion possesses cubic symmetry, which is the upper black dashed
line shown on all figures.
Let us mention that in the region of parameter space to the top-left
of the upper black dashed line, $\sigma^{O_h}$ is very close to zero
everywhere, except close to the dashed line and the deviation here is
merely numerical error.
For $\sigma^{O_h}\gtrsim 0.5$ the loss of octahedral symmetry is
visible to the naked eye, see
figs.~\ref{fig:skarr1}--\ref{fig:skarr3}. 

\begin{figure}[!thp]
\begin{center}
\mbox{\subfloat[]{\includegraphics[width=0.49\linewidth]{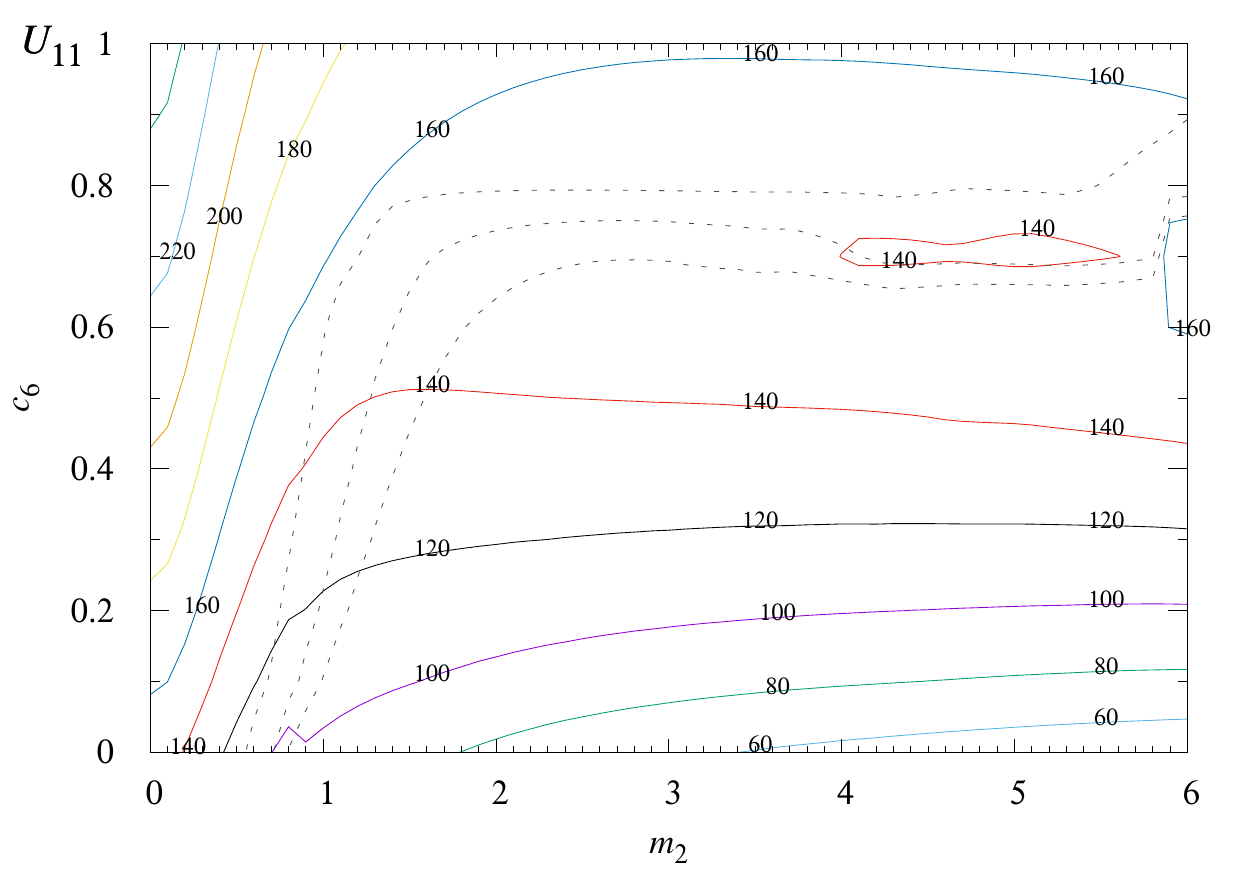}}
\subfloat[]{\includegraphics[width=0.49\linewidth]{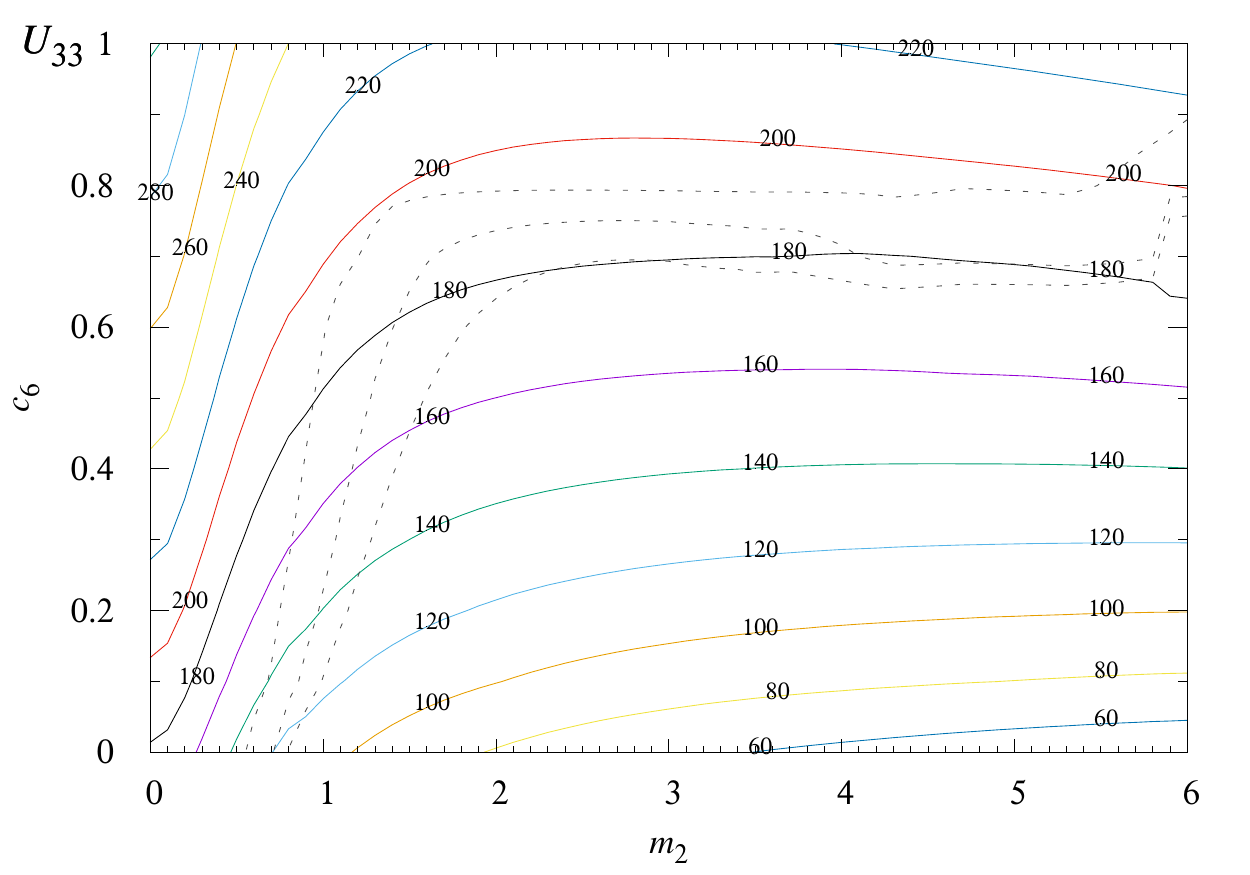}}}
\caption{(a) $U_{11}=U_{22}$ and (b) $U_{33}$ of the isospin inertia
  tensor of the 4-Skyrmion in Skyrme units.
  The dashed lines show contours of $\sigma^{O_h}=0.1,0.5,1$
  from top to bottom. } 
\label{fig:U11U33}
\end{center}
\end{figure}

For completeness we display the values of the inertia tensors in the
parameter space in figs.~\ref{fig:sigmaOh_v}(b) and \ref{fig:U11U33}. 
Throughout the scanned part of the parameter space, we have that
$V_{ij}=v\delta^{ij}$ is diagonal and $W_{ij}=0$ vanishes, whereas the
two nonzero values of the isospin tensor of inertia are
$U_{11}=U_{22}$ and $U_{33}$.
We can see from fig.~\ref{fig:U11U33} that $U_{11}$ is in general
different from $U_{33}$ except in the region of parameter space where
$c_6$ is small and $m_2$ is large.

\section{Discussion}\label{sec:discussion}

In this paper, we have made an extensive parameter scan of the
generalized Skyrme model with the loosely bound potential by
performing full PDE calculations.
The full numerical calculations are necessary for detecting the
symmetries of the Skyrmions, in particular, whether the 4-Skyrmion
possesses the sought-for cubic symmetry or it loses it becoming a
tetrahedrally symmetric object of ``point'' particles.
In ref.~\cite{Gudnason:2016tiz} we made an initial study of the
parameter space limited to $m_2\leq 1$ using the rational map Ansatz
for the cube, under the assumption that the symmetries would stay
unchanged in such a limited part of parameter space. 
That assumption turned out to be true indeed.
In this paper, we have been able to go much further into the direction
of turning up the loosely bound potential and hence reducing the
binding energies significantly.
As expected from ref.~\cite{Gudnason:2016mms}, the loosely bound
potential itself will quickly break the cubic symmetry of the
4-Skyrmion.
Fortunately, it turns out that turning on a finite sextic term makes
the 4-Skyrmion resistant to the impending symmetry breaking a long way
up in $m_2$, the (square-root of the) coefficient of the loosely bound
potential. 

The philosophy that we have used as a guiding principle is to keep as
much symmetry as possible whilst reducing the binding energies as much
as possible; we try to cling on to the platonic symmetries possessed
by the Skyrmions of small $B$ ($B<8$) and in particular the cubic
symmetry of the 4-Skyrmion.
Next we note that the classical relative binding energy (CRBE) is
almost independent of the sextic term with coefficient $c_6$, but
depends strongly on the loosely bound potential with coefficient
$m_2^2$. 
Now as a direct test for whether the above stated philosophy is
justified experimentally, we can compare two limits:
the point particle model limit (PPML) ($c_6=0$ and $m_2\to\infty$) and
the BPS-Skyrme model limit (BSML) ($c_6\propto m_2\to\infty$).
Of course we have not taken any strict limit and just considered $m_2$
large ($m_2\simeq 6$) and the BSML case here will refer to $c_6=1$,
$m_2=6$. 
First of all, we find that although the CRBE is almost the same in the
two cases, the QRBE receives a larger quantum contribution in the PPML
case than in the BSML case. 
This fact is intimately related to the value of the isospin inertia
tensor being larger when the sextic term is turned on, which in turn
is due to the sextic term enlarging the Skyrmions.
The nucleon mass correlates with the binding energy in our calibration
scheme and hence is closer to the measured value in the BSML case (but
still overestimated) than in the PPML case.
On the other hand, the pion decay constant is larger (but still
underestimated) in the PPML case than in the BSML case and the
electric charge radius is almost spot-on the experimental value in the
PPML corner of our parameter space.
Finally there is a tie between the two cases for the axial coupling of
the nucleon, for which it is overestimated in the BSML case and
underestimated in the PPML case.
We present in table \ref{tab:benchmarks} four benchmark points
compared to experimental data, including the PPML and BSML cases.

\begin{table}[!htp]
\begin{center}
\caption{Benchmark points compared to experimental data.
The model points are: A [SSM] Standard Skyrme Model; B [GSM]
Generalized Skyrme Model; C [PPML] Point Particle Model Limit; D
[BSML] BPS-Skyrme Model Limit. }
\label{tab:benchmarks}
\begin{tabular}{l||r@{.}lr@{.}lr@{.}lr@{.}l}
& \multicolumn{2}{c}{Point A [SSM]} & \multicolumn{2}{c}{Point B [GSM]}
& \multicolumn{2}{c}{Point C [PPML]} & \multicolumn{2}{c}{Point D [BSML]}\\
& \multicolumn{2}{c}{$(m_2,c_6)=(0,0)$}
& \multicolumn{2}{c}{$(m_2,c_6)=(0,1)$} & \multicolumn{2}{c}{$(m_2,c_6)=(6,0)$} & \multicolumn{2}{c}{$(m_2,c_6)=(6,1)$}\\
\hline
\hline\\[-12pt]
$\tilde{f}_\pi$ & $-32$&$1\%$ & $-51$&$0\%$ & $-80$&$8\%$ & $-86$&$5\%$\\
$\tilde{m}_N$ & $+12$&$0\%$ & $+13$&$9\%$ & $+6$&$2\%$ & $+2$&$9\%$\\
$\tilde{m}_\Delta$ & $-3$&$8\%$ & $-2$&$6\%$ & $+1$&$6\%$ & $-11$&$0\%$\\
$\tilde{m}_\pi^\pm$ & $+14$&$7\%$ & $+39$&$9\%$ & $-51$&$7\%$ & $-46$&$2\%$\\
$\tilde{m}_\pi^0$ & $+18$&$6\%$ & $+44$&$7\%$ & $-50$&$1\%$ & $-44$&$4\%$\\
$\tilde{g}_A$ & $+17$&$4\%$ & $+30$&$3\%$ & $-30$&$0\%$ & $+13$&$6\%$\\
$\delta_4$ & $+970$&$3\%$ & $+1174$&$5\%$ & $-67$&$7\%$ & $-72$&$5\%$\\
$\delta_4^{\rm tot}$ & $+1331$&$4\%$ & $+1514$&$6\%$ & $+725$&$7\%$ & $+348$&$9\%$\\
$\tilde{r}_{1,E}$ & $+27$&$8\%$ & $+25$&$3\%$ & $-4$&$2\%$ & $+23$&$0\%$
\end{tabular}
\end{center}
\end{table}

For the observables considered in this paper, it is not clear that
preserving as much symmetry as possible is better in line with
phenomenology.
Both the cases summarized above have their merits.
Nevertheless, once the full quantum excitational spectrum is
considered for the nuclei, it becomes clear that having a large
symmetry is not just aesthetics but a necessity.
This was pointed out recently in ref.~\cite{Gudnason:2018aej}; in this
particular case, the lack of a symmetry of rank four resulted in
parity doubling of the states -- not observed in Nature.

A completely different way to try to solve the binding energy problem
of the Skyrmions is to disregard the classical energies completely and
believe that the true quantum states have very large corrections to
their classical counterparts.
This solution may well be the true description of nuclear physics
although it runs against conventional particle physics wisdom that
prefers natural mechanisms as explanations for physical effects.
A known counterexample for naturalness indeed in nuclear physics is
the fact that the binding energy of the triplet deuteron is about 2.2
MeV whereas the energy released in neutron beta decays is about 1.3
MeV. That 0.9 MeV difference is what kept all the neutrons from
decaying during the evolution of the Universe and be missing in the
formation of countless elements.
Of course our Universe may just well be a giant accident. 
As we discussed in the introduction, the question of whether
semi-classical quantization with the perturbative addition of a few
light modes of the soliton is a good approximation comes down to
whether the fluctuation spectrum is ``weakly coupled''.
It would be an important next step to investigate this issue in
depth. 

The $\Delta$ resonance is at best problematic in the Skyrme model.
The reason for this is evident from our discussion about trying to
reduce the quantum isospin contribution to the mass of the nucleon and
is rooted in eq.~\eqref{eq:mDelta}. That is, if we make
$\tilde{\epsilon}_1$ small then so is $5\tilde{\epsilon}_1$, see
ref.~\cite{Adam:2016drk}; more concretely, if we want the spin/isospin 
contribution to the mass of the nucleon, $\tilde{\epsilon}_1$, to be
less than the binding energy of roughly 16 MeV, which is approximately
the binding energy of nuclear matter, then it is impossible for
$5\tilde{\epsilon}_1$ to be as large as 366
MeV \cite{Adam:2016drk}\footnote{In nuclear matter, the rigid body
quantization can be neglected, hence the binding energy of nuclear
matter gives an upper bound on the mass contribution of the spin to
the nucleon, $\tilde{\epsilon}_1\lesssim 16$ MeV.}.
As further pointed out in ref.~\cite{Adam:2016drk}, the $\Delta$
resonance probably needs a fully relativistic treatment and should be
considered as a resonance with a complex mass pole. 

Considering larger coefficients of the sextic BPS-Skyrme term is a
natural continuation of this work and it may show qualitatively
interesting new behavior of the model.
If the BPS-Skyrme term and the potential term become too large,
however, one enters the near-BPS regime of the BPS-Skyrme model which
is known to be technically difficult.
In the light of the discussions in this paper, the more important
question is whether it is necessary to obtain solutions with very low
classical binding energies -- like in this paper -- or the true
solution to the quantum physics of nuclei lies in sizable corrections
that perhaps via beautiful symmetries somehow all balance in such a
way as to give small binding energies at the 1-percent level for all
nuclei. 
This will be left as work for the future.

We should remind the reader that we did not get the physical pion mass
right in the region of parameter space with small binding energies.
This can easily be fixed, but the change for the rest of the physics
is expected to be insignificant and as long as the pion decay constant
is so far from its measured value, it remains a question whether the
pion mass should be close to its experimental value or not.
Lattice-QCD simulations often get good results even though their pion
mass is typically much too large compared with the experimental
value. 

There are lots of directions to consider for improving the Skyrme
model in order for it to become a full-fledged high-precision model
of nuclear physics.
If the program succeeds, it will become a few-parameter model
which basically can cover all nuclei.
The list of problems is however not so short.
The problem of the binding energy that we have worked on in this paper
is not solved yet and we are probably getting closer to a point where
we can determine whether the Skyrme model is natural and hence the
quantum corrections somehow are small as expected in systems with
semi-classical quantization, or there are relatively large quantum
corrections that just happen to balance out almost perfectly over a
large variety of nuclei -- many possessing different symmetries.

The small isospin breaking present in Nature still remains largely
unincorporated in the Skyrme model. The recent suggestion by
Speight \cite{Speight:2018zgc} is based on including the $\omega$
meson and an explicit symmetry breaking term.
This direction of improving the Skyrme model is also considered for
solving the binding energy problem. That is, including vector mesons
in the model, see e.g.~\cite{Naya:2018mpt} where $\rho$ mesons are
considered. 

A further improvement to be considered, which will become more
important for the studies of large nuclei, is to include the effects
of the Coulomb energy. Although it is known how to calculate the
Coulomb force for multi-Skyrmions, it should ideally be back-reacted
onto the Skyrmions. This would require some partial gauging and
further complicate the model. 

It would also be interesting to consider other Skyrmions than the
4-Skyrmion, in order to check our claims about the preservation of
symmetries in the model.
A preliminary study suggests that for the 8-Skyrmion, the two cubes
retain their separate octahedral symmetries but become more weakly
bound to each other with the result that, in the low-binding energy
regime, the chain and twisted chain become almost degenerate in
energies.
There are plenty of other Skyrmions that would be interesting to
study.

The question, however, remains: how to resolve the quantum part of the
binding-energy problem for Skyrmions as nuclei? Hopefully, this
question may be answered in the future.

\subsection*{Acknowledgments}

S.~B.~G.~thanks Chris Halcrow for discussions. 
S.~B.~G.~is supported by the Ministry of Education, Culture, Sports,
Science (MEXT)-Supported Program for the Strategic Research Foundation
at Private Universities ``Topological Science'' (Grant No.~S1511006) and
by a Grant-in-Aid for Scientific Research on Innovative Areas
``Topological Materials Science'' (KAKENHI Grant No.~15H05855) from
MEXT, Japan.
The calculations in this work were carried out using the TSC computing 
cluster of the ``Topological Science'' project at Keio University.

\end{document}